\documentclass[a4paper,12pt]{article}
\usepackage{amsmath,enumerate}
\usepackage{graphicx}
\usepackage{slashed}
\newcommand{\lapprox}{%
\mathrel{%
\setbox0=\hbox{$<$}
\raise0.6ex\copy0\kern-\wd0
\lower0.65ex\hbox{$\sim$}
}}
\newcommand{\gapprox}{%
\mathrel{%
\setbox0=\hbox{$>$}
\raise0.6ex\copy0\kern-\wd0
\lower0.65ex\hbox{$\sim$}
}}

\begin{document}

\begin{center}

{\Large \bf Lepton and quark mixing patterns with generalized $CP$
transformations}\\[20mm]

Joy Ganguly\footnote{ph18resch11009@iith.ac.in} and
Raghavendra Srikanth Hundi\footnote{rshundi@phy.iith.ac.in}\\
Department of Physics, Indian Institute of Technology Hyderabad,\\
Kandi - 502 284, India.\\[20mm]

\end{center}

\begin{abstract}

In this work, we have modified a scenario, originally proposed by Grimus and
Lavoura, in order to obtain maximal values for atmospheric mixing angle and
$CP$ violating Dirac phase of the lepton sector. To achieve this, we have
employed $CP$ and some discrete symmetries in a type II seesaw
model. In order to make predictions about neutrino mass ordering and
the smallness of the reactor angle, we have obtained some conditions on
the elements of the neutrino mass matrix of our model. Finally, within the
framework of our model, we have studied quark masses and mixing pattern.

\end{abstract}

\newpage

\section{Introduction}

From the global fits to neutrino oscillation data \cite{glo-fit} it is known
that the three mixing angles in lepton sector are close to the tribimaximal
(TBM) mixing \cite{tbm}. In the TBM pattern, the three mixing angles take
the following values: $\sin^2\theta_{12}=\frac{1}{3}$,
$\sin^2\theta_{23}=\frac{1}{2}$, $\sin^2\theta_{13}=0$. On the other hand, the
$CP$ violating Dirac phase $\delta_{CP}$ in lepton sector is yet to be measured
precisely. However, from the global fits to neutrino oscillation data
\cite{glo-fit}, the
best fit value for $\delta_{CP}$ is around $\pi(\frac{3}{2}\pi)$ in the case of
normal(inverted) ordering of neutrino masses. The TBM value for $\theta_{23}$
and $\delta_{CP}=\frac{3}{2}\pi$ are still allowed in the $3\sigma$ ranges
for these observables in the current neutrino oscillation data \cite{glo-fit}.
The above mentioned values for $\theta_{23}$ and $\delta_{CP}$ are considered
to be maximal. To explain the maximal values for $\theta_{23}$ and
$\delta_{CP}$,
Harrison and Scott have proposed $\mu-\tau$ symmetry in
combination with $CP$ symmetry, which together is called $\mu-\tau$
reflection symmetry \cite{hs}. For some works based on $\mu-\tau$ symmetry
and $CP$ symmetry, see Refs. \cite{rev,mutau}. Ref. \cite{rev}
is a review article.

In the work by Grimus and Lavoura \cite{gl}, it is shown that a mass matrix
for light left-handed neutrinos of the following form \cite{bmv}
\begin{equation}
{\cal M}_\nu=\left(\begin{array}{ccc}
a & r & r^* \\ r & s & b \\ r^* & b & s^*
\end{array}\right)
\label{eq:gl}
\end{equation}
can yield maximal values for $\theta_{23}$ and $\delta_{CP}$. In the above
equation, $a,b$ are real and $r,s$ are complex. Further, in Ref.
\cite{gl}, a model is constructed, which is based on $\mu-\tau$ reflection
symmetry and softly broken lepton numbers,
in order to obtain a mass matrix of the form of Eq. (\ref{eq:gl}) for
light neutrinos. In this model, three Higgs doublets are introduced and
light neutrinos acquire masses via type I seesaw mechanism \cite{t1s}. Lepton
number is softly broken by the mass terms for right-handed neutrinos of
this model. In this model, in the absence of fine tuning of the parameters, muon
and tau leptons can have masses of the same order. To explain the hierarchy
in the masses for
these leptons, $K$ symmetry is introduced, under which the muon is massless
\cite{gl}. Realistic masses for muon and
tau leptons are explained in the above mentioned scenario with the soft
breaking of the $K$ symmetry \cite{gl2}. The work done together in
Refs. \cite{gl,gl2}, which is based on $\mu-\tau$ reflection symmetry,
consistently explains the mixing pattern in lepton sector
and also the masses for charged leptons.

Although the work done in Refs. \cite{gl,gl2}
gives a consistent picture about masses and mixing pattern in the lepton
sector, there are few limitations for this work which are explained
below. It is argued in Ref. \cite{gl} that the mass matrix of Eq.
(\ref{eq:gl}), which is obtained from $\mu-\tau$ reflection symmetry, cannot
give predictions about neutrino mass ordering and the mixing angle
$\theta_{12}$. Moreover, in the case that the $\delta_{CP}$ is maximal, the mass
matrix of Eq. (\ref{eq:gl}) cannot predict anything about $\theta_{13}$
\cite{gl}. From the current neutrino oscillation data it is known that
neutrinos can have either normal or inverted mass ordering, and moreover,
we have $\sin^2\theta_{12}\sim\frac{1}{3}$ and
$\sin^2\theta_{13}\sim10^{-2}$ \cite{glo-fit}. Apart from the above
mentioned limitations, in Ref. \cite{gl}, mixing pattern in the quark sector
is not addressed. It is known that mixing pattern in the quark sector
\cite{pdg} is quite different from that of lepton sector. One would like
to know if the mixing patterns for both quark and lepton sectors can be
understood in the same framework.

As stated above, in Ref. \cite{gl}, a model, which is based on type I seesaw
mechanism
and $\mu-\tau$ reflection symmetry, is presented in order to obtain the
neutrino mass matrix of the form of Eq. (\ref{eq:gl}). In this work, our aim
to see if the matrix of Eq. (\ref{eq:gl}) can be obtained with type II seesaw
mechanism \cite{t2s} in the framework of $\mu-\tau$ reflection symmetry.
In order to achieve this, we construct a model which has three Higgs doublets
and one scalar Higgs triplet. In our model, right-handed neutrinos do not
exist, and hence, neutrinos acquire masses when the Higgs triplet get vacuum
expectation value (VEV). The purpose of Higgs doublets is to give masses to
charged leptons via Yukawa couplings. With the $\mu-\tau$ reflection symmetry
in our model, if the
VEV of Higgs triplet is real, we show that the light neutrinos
will have the mass matrix of the form of Eq. (\ref{eq:gl}). In order to show
if the VEV of Higgs triplet can be real, we analyze the scalar potential
of our model. We demonstrate that by using an extra discrete symmetry, the
VEV of Higgs triplet can be real. While analyzing the scalar potential of our
model, we also address the problem of hierarchy in the masses of muon and
tau leptons. In the literature, models have been constructed in order to achieve
maximal values for $\theta_{23}$ and $\delta_{CP}$ using type II
seesaw mechanism
\cite{gl3,maho-nis}. However, in these models, multiple Higgs triplets have
been introduced in addition to the three Higgs doublets. Hence, the model
we have proposed here is economical as compared to the above mentioned models.

As stated previously, the form of mass matrix given in Eq. (\ref{eq:gl})
can predict maximal values for $\theta_{23}$ and
$\delta_{CP}$. But this matrix cannot give predictions about neutrino mass
ordering and the mixing
angles $\theta_{12},\theta_{13}$. As already pointed before, we have
$\sin^2\theta_{13}\sim10^{-2}$ \cite{glo-fit}, which means $\theta_{13}$
is a small angle. In this work, we do an analysis, based on some approximation
procedure \cite{prwo}, and derive some conditions on the elements of neutrino
mass matrix which can predict about neutrino mass ordering and the smallness
of $\theta_{13}$, apart from giving
maximal values for $\theta_{23}$ and $\delta_{CP}$. In this analysis, we assume
$\sin^2\theta_{12}\sim\frac{1}{3}$. To achieve the above mentioned conditions,
new mechanisms should be proposed. In this work, we have attempted to give
one mechanism in order to achieve one of those conditions.

Since in our model three Higgs doublets exist, which give masses to charged
leptons, it is worth to know if these scalar doublets can also generate masses
and mixing pattern for quarks. Due to $CP$ symmetry in the lepton sector,
it is found that these Higgs doublets should transform non-trivially under the
$CP$ symmetry. As a result of this, we propose $CP$ transformations for quarks
in such a way that the Yukawa couplings for these are invariant under the $CP$
symmetry.
It is known that there is a large hierarchy among the masses of quarks.
Hence, in order to explain the mixing pattern for quarks, the Yukawa couplings
for these should be hierarchically suppressed \cite{rasin}. To explain the
realistic mixing pattern for quarks through hierarchically suppressed
Yukawa couplings, we have followed the work done in Refs. \cite{bn,lmn}.
To know about other works
on quark and lepton mixings with generalized $CP$ transformations,
see Ref. \cite{owql}. There are some works which addressed quark and
lepton mixings with other symmetries. For example, see Refs. \cite{rr}.

The paper is organized as follows. In the next section, we propose a model
for lepton mixing, where maximal values for $\theta_{23}$ and $\delta_{CP}$
can be predicted if the VEV of triplet Higgs is real. In Sec. 3, we analyze
the scalar potential of our model and show that the VEV of triplet Higgs
can be real if we introduce additional discrete symmetry $Z_3$. In Sec. 4,
we obtain some conditions on the elements of the neutrino mass matrix of
our model, which can predict about the neutrino mass ordering and the
smallness of $\theta_{13}$. In Sec. 5, we study on the quark masses and mixing
pattern and demonstrate that they can be explained in the framework of our
model. We conclude in the last section. In the Appendix, we attempt to give
a mechanism for achieving normal order of neutrino masses.

\section{A model for lepton mixing}

The model we propose for lepton mixing is similar to that in Ref. \cite{gl}.
We propose scalar Higgs doublets $\phi_i=(\phi_i^+,\phi_i^0)^T$, where
$i=1,2,3$, in order to give masses
to charged leptons. We denote the lepton doublets and singlets by
$D_{\alpha L}=(\nu_{\alpha L},\alpha_L)^T$ and $\alpha_R$, where
$\alpha=e,\mu,\tau$, respectively.
The $CP$ transformations on the lepton fields and Higgs doublets are defined
as \cite{gl}
\begin{eqnarray}
&& D_{\alpha L}\to iS_{\alpha\beta}\gamma^0C\bar{D}_{\beta L}^T,\quad
\alpha_R\to iS_{\alpha\beta}\gamma^0C\bar{\beta}_R^T,
\quad S=\left(\begin{array}{ccc} 1 & 0 & 0 \\ 0 & 0 & 1 \\ 0 & 1 & 0 \end{array}
\right),
\nonumber \\
&& \phi_{1,2}\to\phi_{1,2}^*,\quad \phi_3\to -\phi_3^*.
\end{eqnarray}
Here, $C$ is the charge conjugation matrix.
In addition to the invariance under the above mentioned $CP$ transformations,
one needs to impose conservation
of $U(1)_{L_\alpha}$ and $Z_2$ symmetries. Here, $U(1)_{L_\alpha}$ is the
lepton number symmetry for the individual family of leptons. Under $Z_2$
symmetry, only the $e_R$ and $\phi_1$ change sign.
With the above mentioned
charge assignments, the invariant Lagrangian for charged lepton Yukawa
couplings is given by \cite{gl}
\begin{equation}
{\cal L}_Y = -y_e\bar{D}_{eL}\phi_1e_R-\sum_{j=2}^3\sum_{\alpha=\mu,\tau}
g_{j\alpha}\bar{D}_{\alpha L}\phi_j\alpha_R +h.c..
\label{eq:clep}
\end{equation}

In order for the Lagrangian of Eq. (\ref{eq:clep}) to be invariant under $CP$
symmetry, we should have $y_e$
to be real, $g_{2\mu}=g_{2\tau}^*$ and $g_{3\mu}=-g_{3\tau}^*$. Since the
mass of electron should be real, we take VEV of $\phi_1$ to be real. On
the other hand, the VEVs of $\phi_{2,3}$ should be complex, which give
masses to muon and tau leptons, whose forms are given below \cite{gl}.
\begin{equation}
m_\mu=|g_{2\mu}v_2+g_{3\mu}v_3|,\quad
m_\tau=|g^*_{2\mu}v_2-g^*_{3\mu}v_3|.
\label{eq:muntau}
\end{equation}
In this work, we are taking $\langle\phi_i^0\rangle=v_i$ for $i=1,2,3$.
A priori, the VEVs of all Higgs doublets are of the same order. Hence, from
the above equations, we can notice that some fine tuning is necessary in order
to explain the hierarchy in the muon and tau lepton masses. To reduce this
fine tuning, $K$ symmetry is introduced, under which the non-trivial
transformations of the fields are given below \cite{gl}
\begin{equation}
\mu_R\to -\mu_R,\quad \phi_2\leftrightarrow\phi_3.
\end{equation}
After imposing this $K$ symmetry in the above described model, one can see
that $g_{2\mu}=-g_{3\mu}$. Using this in Eq. (\ref{eq:muntau}), we get
\begin{equation}
\frac{m_\mu}{m_\tau}=\left|\frac{v_2-v_3}{v_2+v_3}\right|.
\label{eq:mutau}
\end{equation}
Since the scalar potential of this model should also respect the $K$ symmetry,
we should get $v_2=v_3$, and hence, $m_\mu=0$. Now, to explain a non-zero but
small $m_\mu$, soft breaking of $K$ symmetry can be introduced into the scalar
potential of this model \cite{gl2}. Analysis related to this is presented
in the next section.

To explain the masses for neutrinos in the above described framework, we
introduce the following Higgs triplet into the model.
\begin{equation}
\Delta=\left(\begin{array}{cc}
\frac{\Delta^+}{\sqrt{2}} & \Delta^{++} \\
-\Delta^0 & -\frac{\Delta^+}{\sqrt{2}}
\end{array}\right).
\end{equation}
$\Delta$ is singlet under $Z_2$, but otherwise transform under $CP$ symmetry as
$\Delta\to\Delta^*$. Now, the Yukawa couplings for neutrinos can be written as
\begin{equation}
{\cal L}_Y=\frac{1}{2}\sum_{\alpha,\beta=e,\mu,\tau}Y^\nu_{\alpha\beta}
\bar{D}^c_{\alpha L}i\sigma_2\Delta D_{\beta L}+h.c..
\label{eq:lagnu}
\end{equation}
Here, $D_{\alpha L}^c$ is the charge conjugated doublet for $D_{\alpha L}$
and $\sigma_2$ is a Pauli matrix. We can notice that the terms in the above
Lagrangian break the lepton number symmetry $U_{L_\alpha}$ explicitly. We
can consider this technically natural, since in the limit that the symmetry
$U_{L_\alpha}$ is exact, the neutrino masses become zero in this model. Hence
to explain the smallness of neutrino masses we can break $U_{L_\alpha}$
symmetry by a small amount. As a result
of this, the neutrino Yukawa couplings $Y^\nu_{\alpha\beta}$ can be small in
this work. Due to the invariance under $CP$ symmetry, these Yukawa couplings
should satisfy
\begin{equation}
SY^\nu S=(Y^\nu)^*.
\label{eq:ynu}
\end{equation}
After electroweak symmetry breaking, we can have
$\langle\Delta^0\rangle=v_\Delta$. Now, from Eq. (\ref{eq:lagnu}), we get the
mass matrix for neutrinos, which is given by $M_\nu=Y^\nu v_\Delta$.
If $v_\Delta$ is real, using Eq. (\ref{eq:ynu}), we get
\begin{equation}
SM_\nu S=M_\nu^*.
\label{eq:mnu}
\end{equation}
In order to satisfy the above relation, the form for $M_\nu$ should be same
as that of Eq. (\ref{eq:gl}). Hence, in the above proposed model, the mixing
angle $\theta_{23}$ and the $CP$ violating phase $\delta_{CP}$ are maximal.
However,
in order to satisfy the relation in Eq. (\ref{eq:mnu}), $v_\Delta$ should
be real. In the next section, we present an analysis of scalar potential
of our model,
where we demonstrate that $v_\Delta$ can be real.

Let us get some estimation on the value of $v_\Delta$ in our work. As stated
above, the neutrino mass matrix in our work is $M_\nu=Y^\nu v_\Delta$. Since
the couplings $Y^\nu$ need to small, because they break the $U_{L_\alpha}$
symmetry by a small amount, we take $Y^\nu\sim 10^{-3}$. Now, by fitting
$M_\nu$ to
neutrino masses, which are obtained from neutrino oscillation data, we can get
some estimation about $v_\Delta$. Using
the neutrino oscillation data, the following mass-square
differences have been found \cite{glo-fit}, where we have given the best
fit values.
\begin{equation}
m_s^2\equiv m_2^2-m_1^2=7.5\times10^{-5}~{\rm eV}^2,\quad
m_a^2\equiv\left\{\begin{array}{c}
m_3^2-m_1^2=2.55\times10^{-3}~{\rm eV}^2~~{\rm (NO)}\\
m_1^2-m_3^2=2.45\times10^{-3}~{\rm eV}^2~~{\rm (IO)}
\end{array}\right..
\label{eq:msquare}
\end{equation}
Here, $m_{1,2,3}$ are neutrino mass eigenvalues and NO(IO) represents
normal(inverted) ordering. Using the above values, we get $m_s\sim0.0087$ eV
and $m_a\sim0.05$ eV, which correspond to solar and atmospheric neutrino
mass scales, respectively. In order to fit these neutrino mass scales in our
work, we can take $v_\Delta\sim1-10$ eV.

\section{Analysis of scalar potential}

The scalar fields of the model proposed in the previous section are charged
under the symmetry $CP\times Z_2\times K$. The invariant scalar potential of
this model can be written as
\begin{equation}
V_{inv}=V_D+V_T.
\label{eq:vin}
\end{equation}
Here, $V_D$ contains potential terms only for the Higgs doublets. $V_T$
is the scalar potential for the triplet Higgs of our model.
The form of $V_D$ is given by \cite{gl2}
\begin{eqnarray}
V_D&=& -M_1^2\phi_1^\dagger\phi_1
-M_2^2(\phi_2^\dagger\phi_2+\phi_3^\dagger\phi_3)
+\lambda_1(\phi_1^\dagger\phi_1)^2
+\lambda_2\left[(\phi_2^\dagger\phi_2)^2+(\phi_3^\dagger\phi_3)^2\right]
\nonumber \\
&&+\lambda_3(\phi_1^\dagger\phi_1)(\phi_2^\dagger\phi_2+\phi_3^\dagger\phi_3)
+\lambda_4(\phi_2^\dagger\phi_2)(\phi_3^\dagger\phi_3)
+\lambda_5\left[(\phi_1^\dagger\phi_2)(\phi_2^\dagger\phi_1)
+(\phi_1^\dagger\phi_3)(\phi_3^\dagger\phi_1)\right]
\nonumber \\
&&+\lambda_6\left[(\phi_2^\dagger\phi_3)(\phi_3^\dagger\phi_2)\right]
+\lambda_7\left[(\phi_2^\dagger\phi_3)^2+(\phi_3^\dagger\phi_2)^2\right]
\nonumber \\
&&+\lambda_8\left[(\phi_1^\dagger\phi_2)^2+(\phi_1^\dagger\phi_3)^2
+(\phi_2^\dagger\phi_1)^2+(\phi_3^\dagger\phi_1)^2\right]
\nonumber \\
&&+i\lambda_9\left[(\phi_1^\dagger\phi_2)(\phi_1^\dagger\phi_3)
-(\phi_2^\dagger\phi_1)(\phi_3^\dagger\phi_1)\right]
+i\lambda_{10}(\phi_2^\dagger\phi_3-\phi_3^\dagger\phi_2)
(\phi_2^\dagger\phi_2-\phi_3^\dagger\phi_3).
\label{eq:vd}
\end{eqnarray}
In the above equation, all parameters are real due to either hermiticity
or $CP$ symmetry of the potential. To obtain
$V_T$, we have followed the work of Ref. \cite{maet}. The form of $V_T$ is
given below.
\begin{eqnarray}
V_T&=& m_\Delta^2{\rm Tr}(\Delta^\dagger\Delta)
+\frac{1}{2}\lambda_\Delta[{\rm Tr}(\Delta^\dagger\Delta)]^2
+\lambda_{11}\phi_1^\dagger\phi_1{\rm Tr}(\Delta^\dagger\Delta)
+\lambda_{12}(\phi_2^\dagger\phi_2+\phi_3^\dagger\phi_3)
{\rm Tr}(\Delta^\dagger\Delta)
\nonumber \\
&&+\lambda_{13}{\rm Tr}(\Delta^\dagger\Delta^\dagger){\rm Tr}(\Delta\Delta)
+\lambda_{14}\phi_1^\dagger\Delta^\dagger\Delta\phi_1
+\lambda_{15}(\phi_2^\dagger\Delta^\dagger\Delta\phi_2
+\phi_3^\dagger\Delta^\dagger\Delta\phi_3)
\nonumber \\
&&+\kappa_1(\tilde{\phi}_1^Ti\sigma_2\Delta\tilde{\phi}_1+h.c.)
+\kappa_2(\tilde{\phi}_2^Ti\sigma_2\Delta\tilde{\phi}_2
+\tilde{\phi}_3^Ti\sigma_2\Delta\tilde{\phi}_3+h.c.)
\nonumber \\
&&+i\kappa_3(\tilde{\phi}_2^Ti\sigma_2\Delta\tilde{\phi}_3-h.c.)
\label{eq:vt}
\end{eqnarray}
Here, $\tilde{\phi}_k=i\sigma_2\phi_k^*,k=1,2,3$.
All the parameters in the above equation are real, due to either hermiticity
or $CP$ symmetry of the potential.

As described in the previous section, the VEV of $\phi_1$ is real but
the VEVs for $\phi_{2,3}$ should be complex. Although all parameters in
Eq. (\ref{eq:vt}) are real, due to complex VEVs of $\phi_{2,3}$, the trilinear
terms containing $\kappa_{2,3}$ can contribute complex VEV to $\Delta$.
However, it
may happen that the phases of the VEVs of $\phi_{2,3}$ can be fine tuned
in such a way that the $\kappa_{2,3}$-terms can give a real VEV to $\Delta$.
We study these points by minimizing the scalar potential of our model. Before
we present that study, we have to estimate the order of magnitudes of the
unknown parameters of Eqs. (\ref{eq:vd}) and (\ref{eq:vt}). From the
naturalness argument, we take all dimensionless $\lambda$ parameters to be
${\cal O}(1)$. Since the VEVs of Higgs doublets should be around the electroweak
scale of $v_{EW}=$ 174 GeV, we take $M_1^2,M_2^2\sim v_{EW}^2$. Now, we have
to determine the order of magnitudes for $m_\Delta^2$ and $\kappa_{1,2,3}$. This
is explained below. After minimizing the potential of Eq. (\ref{eq:vt}) with
respect to $\Delta^0$, we naively get
\begin{equation}
v_\Delta\sim\frac{\kappa v_{EW}^2}{m_\Delta^2+\lambda v_{EW}^2}.
\end{equation}
Here, $\kappa\sim\kappa_{1,2,3}$ and $\lambda\sim\lambda_{11,12}$. While
obtaining
the above equation, we have used $v_\Delta\ll v_{EW}$. In order to get a
very small $v_\Delta$, we can consider the following two cases.
\begin{eqnarray}
&& {\rm case~I}:\quad m_\Delta\gg v_{EW},\quad \kappa\sim m_\Delta.
\nonumber \\
&& {\rm case~II}:\quad m_\Delta\sim v_{EW},\quad \kappa\sim v_\Delta.
\end{eqnarray}
In case I, the smallness of $v_\Delta$ is explained by taking a large value
for $m_\Delta$, which is around $10^{12}$ GeV. In case II, by suppressing
the $\kappa$ parameters, one can understand the smallness of $v_\Delta$. In
case I, the value of $m_\Delta$ is close to the breaking scale of supersymmetry
in supergravity models \cite{susy}. Hence, one can motivate case I from
supersymmetry. On the other hand, in case II, one has to find a mechanism
for the suppression of $\kappa$ parameters. From the phenomenology point of
view, case II can be tested in the LHC experiment, since the masses for
the components of scalar triplet Higgs can be around few 100 GeV.

In case I, we can notice that $\langle V_D\rangle\sim\langle V_T\rangle$.
Only the terms containing $m_\Delta^2$ and $\kappa$ parameters in
$\langle V_T\rangle$ can be of the order of $\langle V_D\rangle$. Other
terms in $\langle V_T\rangle$ give negligibly small contribution in
comparison to $\langle V_D\rangle$. On the other hand, in case II,
$\langle V_T\rangle\ll\langle V_D\rangle$. Because of this difference in the
contribution of $V_T$ in both these cases, we minimize the scalar potential
of our model separately for these two cases.

\subsection{Case I}

We parametrize the VEVs of scalar fields as follows.
\begin{equation}
\langle\phi_1^0\rangle=v_1,\quad
\langle\phi_2^0\rangle=v_2=v\cos\sigma e^{i\alpha},\quad
\langle\phi_3^0\rangle=v_3=v\sin\sigma e^{i\beta},\quad
\langle\Delta^0\rangle=v_\Delta=v^\prime e^{i\theta}.
\label{eq:para}
\end{equation}
Here, $v_1,v,v^\prime$ are real.
We plug the above parametrizations in the scalar potential of Eq.
(\ref{eq:vin}). Since we want $\langle\Delta^0\rangle$ to be real, and
moreover, $V_{inv}$ respect $K$ symmetry, we look for a minimum at
\begin{equation}
\sigma=\frac{\pi}{4},\quad\alpha=\beta=\omega,\quad\theta=0.
\label{eq:min}
\end{equation}
Now, we take first derivatives of $V_{inv}$ with respect to
$\sigma,\alpha,\beta,\theta$ at the values mentioned in Eq. (\ref{eq:min}).
Thereafter, we get the following two conditions.
\begin{eqnarray}
&& 2\lambda_8\sin2\omega+\lambda_9\cos2\omega=0,
\label{eq:dcon}
\\
&& 2\kappa_2\sin2\omega=\kappa_3\cos2\omega.
\label{eq:tcon}
\end{eqnarray}
By satisfying the above two conditions, Eq. (\ref{eq:min}) gives a minimum
to our scalar potential. We justify that this a minimum, after computing
the second derivatives of the potential. This analysis is presented shortly
later. However, with the minimum of Eq. (\ref{eq:min}), we get $v_2=v_3$.
Hence, $m_\mu=0$, which follows from Eq. (\ref{eq:mutau}). To get non-zero
and small $m_\mu$, one should add $K$-violating terms in our model, which
break $K$ symmetry explicitly by a small amount. Here we can see the analogy
between $U_{L_\alpha}$ and $K$ symmetries of our model. Both of these symmetries
are broken explicitly by a small amount in order to generate small masses for
neutrinos and muon.

After including the $K$-violating terms, the procedure we follow for
minimization of the scalar potential is similar to what it is done in
Ref. \cite{gl2}. However, in this work, we write more general form for
$K$-violating terms as compared to that in Ref. \cite{gl2}. In Ref. \cite{gl2},
only the soft terms which break the $K$ symmetry are considered.
The general form for $K$-violating terms in our model, which
respect the symmetry $CP\times Z_2$, is given by
\begin{eqnarray}
V_{\slashed{K}}&=&i\delta M_s^2(\phi_2^\dagger\phi_3-\phi_3^\dagger\phi_2)
+\delta M_2^2\phi_2^\dagger\phi_2+\delta M_3^2\phi_3^\dagger\phi_3
\nonumber \\
&&+\delta\kappa_2(\tilde{\phi}_2^Ti\sigma_2\Delta\tilde{\phi}_2+h.c.)
+\delta\kappa_2^\prime(\tilde{\phi}_3^Ti\sigma_2\Delta\tilde{\phi}_3+h.c.)
\nonumber \\
&&+\delta\lambda_2(\phi_2^\dagger\phi_2)^2
+\delta\lambda_2^\prime(\phi_3^\dagger\phi_3)^2
+\delta\lambda_3(\phi_1^\dagger\phi_1)(\phi_2^\dagger\phi_2)
+\delta\lambda_3^\prime(\phi_1^\dagger\phi_1)(\phi_3^\dagger\phi_3)
\nonumber \\
&&+\delta\lambda_5(\phi_1^\dagger\phi_2)(\phi_2^\dagger\phi_1)
+\delta\lambda_5^\prime(\phi_1^\dagger\phi_3)(\phi_3^\dagger\phi_1)
\nonumber \\
&&+\delta\lambda_8\left[(\phi_1^\dagger\phi_2)^2+(\phi_2^\dagger\phi_1)^2\right]
+\delta\lambda_8^\prime\left[(\phi_1^\dagger\phi_3)^2
+(\phi_3^\dagger\phi_1)^2\right]
\nonumber \\
&&+i\delta\lambda_{10}(\phi_2^\dagger\phi_2)(\phi_2^\dagger\phi_3
-\phi_3^\dagger\phi_2)
+i\delta\lambda_{10}^\prime(\phi_3^\dagger\phi_3)(\phi_2^\dagger\phi_3
-\phi_3^\dagger\phi_2)
\nonumber \\
&&+i\delta\lambda_s(\phi_1^\dagger\phi_1)(\phi_2^\dagger\phi_3
-\phi_3^\dagger\phi_2)
+i\delta\lambda_s^\prime\left[(\phi_1^\dagger\phi_2)(\phi_3^\dagger\phi_1)
-(\phi_2^\dagger\phi_1)(\phi_1^\dagger\phi_3)\right]
\nonumber \\
&&+\delta\lambda_{12}\phi_2^\dagger\phi_2{\rm Tr}(\Delta^\dagger\Delta)
+\delta\lambda_{12}^\prime\phi_3^\dagger\phi_3{\rm Tr}(\Delta^\dagger\Delta)
+\delta\lambda_{15}\phi_2^\dagger\Delta^\dagger\Delta\phi_2
+\delta\lambda_{15}^\prime\phi_3^\dagger\Delta^\dagger\Delta\phi_3
\nonumber \\
&&+i\delta\lambda_t(\phi_2^\dagger\phi_3-\phi_3^\dagger\phi_2)
{\rm Tr}(\Delta^\dagger\Delta)
+i\delta\lambda_t^\prime(\phi_2^\dagger\Delta^\dagger\Delta\phi_3
-\phi_3^\dagger\Delta^\dagger\Delta\phi_2).
\label{eq:soft}
\end{eqnarray}
All parameters in the above equation are real, due to either hermiticity or
$CP$ symmetry of the potential. Terms in the first and
second lines are quadratic and trilinear, respectively. Rest of the terms
in the above equation are quartic.

In Ref. \cite{gl2}, only the soft terms
which are quadratic are given. Moreover, the last two terms in the first line
of Eq. (\ref{eq:soft}) are given in Ref. \cite{gl2}, but by taking
$\delta M_2^2=-\delta M_3^2$. We can notice that if $\delta M_2^2
=\delta M_3^2$, sum of the corresponding terms in Eq. (\ref{eq:soft})
is $K$-symmetric. Hence, as long as $\delta M_2^2\neq\delta M_3^2$, each of
these corresponding terms in Eq. (\ref{eq:soft}) is $K$-violating
but conserve $CP\times Z_2$. Based on this observation, we have constructed
other $K$-violating terms in Eq. (\ref{eq:soft}). Since the terms in
Eq. (\ref{eq:soft}) break $K$ symmetry by a small amount, the parameters for
these should be small as compared to the corresponding parameters of $V_{inv}$.

After including the $K$-violating terms, the total scalar potential of
our model is
\begin{equation}
V_{total}=V_{inv}+V_{\slashed{K}}.
\end{equation}
Previously, we minimized $V_{inv}$ and argued that the minimum can be at
Eq. (\ref{eq:min}). Now, due to the presence of $V_{\slashed{K}}$, the above minimum
can be shifted by a small amount. Due to this, the minimum for $V_{total}$
in terms of small deviations $\delta_0,\delta_+,\delta_-,\delta_\theta$ can
be written as
\begin{equation}
\sigma=\frac{\pi}{4}-\frac{\delta_0}{2},\quad
\alpha=\omega+\delta_++\frac{\delta_-}{2},\quad
\beta=\omega+\delta_+-\frac{\delta_-}{2},\quad
\theta=0+\delta_\theta.
\label{eq:mindev}
\end{equation}
Now, we express $\langle V_{inv}\rangle$ and $\langle V_{\slashed{K}}\rangle$ as a
series summation up to second and first order, respectively, in the above
mentioned small deviations. After neglecting the constant terms, we get
\begin{equation}
\langle V_{total}\rangle=\frac{1}{2}\sum_{a,b}{\cal F}_{ab}\delta_a\delta_b
+\sum_{a}f_a\delta_a.
\label{eq:vdev}
\end{equation}
Here, ${\cal F}_{ab}$ is symmetric in the indices $a,b$ and it corresponds to
second derivatives of $V_{inv}$ calculated at Eq. (\ref{eq:min}). Non-vanishing
elements of ${\cal F}_{ab}$ are given below.
\begin{eqnarray}
&&{\cal F}_{++}=(-8\lambda_8\cos2\omega+4\lambda_9\sin2\omega)v_1^2v^2
+(8\kappa_2\cos2\omega+4\kappa_3\sin2\omega)v^2v^\prime,
\nonumber \\
&&{\cal F}_{--}=-2\lambda_7v^4-2\lambda_8v_1^2v^2\cos2\omega
+2\kappa_2v^2v^\prime\cos2\omega,
\nonumber \\
&&{\cal F}_{00}=-\frac{1}{2}\tilde{\lambda}v^4
+(\lambda_9v_1^2+\kappa_3v^\prime)v^2\sin2\omega,
\nonumber \\
&&{\cal F}_{\theta\theta}=2\kappa_1v_1^2v^\prime+2\kappa_2v^2v^\prime
\cos2\omega+\kappa_3v^2v^\prime\sin2\omega,
\nonumber \\
&&{\cal F}_{0-}=-2\lambda_8v_1^2v^2\sin2\omega+\lambda_{10}v^4
+2\kappa_2v^2v^\prime\sin2\omega,
\nonumber \\
&&{\cal F}_{+\theta}=-2(2\kappa_2\cos2\omega+\kappa_3\sin2\omega)v^2v^\prime.
\label{eq:F}
\end{eqnarray}
Here, $\tilde{\lambda}=-2\lambda_2+\lambda_4+\lambda_6+2\lambda_7$.
The expressions for $f_a$ are given below.
\begin{eqnarray}
&&f_0=\frac{1}{2}(\delta M_2^2-\delta M_3^2)v^2
-(\delta\kappa_2-\delta\kappa_2^\prime)v^2v^\prime\cos2\omega
+\frac{1}{2}(\delta\lambda_2-\delta\lambda_2^\prime)v^4
\nonumber \\
&&+\frac{1}{2}[\delta\lambda_3-\delta\lambda_3^\prime
+\delta\lambda_5-\delta\lambda_5^\prime
+2(\delta\lambda_8-\delta\lambda_8^\prime)\cos2\omega](v_1v)^2
+\frac{1}{2}(\delta\lambda_{12}-\delta\lambda_{12}^\prime)(vv^\prime)^2,
\nonumber \\
&&f_-=\delta M_s^2v^2+(\delta\kappa_2-\delta\kappa_2^\prime)
v^2v^\prime\sin2\omega
-(\delta\lambda_8-\delta\lambda_8^\prime)\sin2\omega(v_1v)^2
+\frac{1}{2}(\delta\lambda_{10}+\delta\lambda_{10}^\prime)v^4
\nonumber \\
&&+(\delta\lambda_s-\delta\lambda_s^\prime)(v_1v)^2
+\delta\lambda_t(vv^\prime)^2,
\nonumber \\
&&f_+=2(\delta\kappa_2+\delta\kappa_2^\prime)v^2v^\prime\sin2\omega
-2(\delta\lambda_8+\delta\lambda_8^\prime)\sin2\omega(v_1v)^2,
\nonumber \\
&&f_\theta=-(\delta\kappa_2+\delta\kappa_2^\prime)v^2v^\prime\sin2\omega.
\label{eq:f}
\end{eqnarray}
Using Eq. (\ref{eq:vdev}), the small deviations in the minimum of $V_{total}$
can be obtained as
\begin{equation}
\delta=-{\cal F}^{-1}f.
\label{eq:dev}
\end{equation}
Here, $\delta=(\delta_0,\delta_-,\delta_+,\delta_\theta)^T$,
$f=(f_0,f_-,f_+,f_\theta)^T$ and ${\cal F}$ is a matrix containing the
elements ${\cal F}_{ab}$.

Since some elements of ${\cal F}_{ab}$ are zero,
Eq. (\ref{eq:dev}) can be decomposed into
\begin{eqnarray}
&&\left(\begin{array}{c}\delta_0 \\ \delta_-\end{array}\right)
=-{\cal F}_1^{-1}\left(\begin{array}{c}f_0 \\ f_-\end{array}\right),\quad
{\cal F}_1=\left(\begin{array}{cc}
{\cal F}_{00} & {\cal F}_{0-} \\ {\cal F}_{0-} & {\cal F}_{--}
\end{array}\right),
\nonumber \\
&&\left(\begin{array}{c}\delta_+ \\ \delta_\theta\end{array}\right)
=-{\cal F}_2^{-1}\left(\begin{array}{c}f_+ \\ f_\theta\end{array}\right),\quad
{\cal F}_2=\left(\begin{array}{cc}
{\cal F}_{++} & {\cal F}_{+\theta} \\ {\cal F}_{+\theta} &
{\cal F}_{\theta\theta}\end{array}\right).
\label{eq:devexp}
\end{eqnarray}
We can see that ${\cal F}$ is in block diagonal form containing ${\cal F}_1$
and ${\cal F}_2$. As stated before, the elements of ${\cal F}$ correspond to
second derivatives of $V_{inv}$ calculated at Eq. (\ref{eq:min}). As a result
of this, if the
eigenvalues of ${\cal F}_1$ and ${\cal F}_2$ are positive then
Eq. (\ref{eq:min}) give minimum to the scalar potential in the absence of
$V_{\slashed{K}}$.
One can see that the unknown $\lambda$ and $\kappa$ parameters of
${\cal F}_{1,2}$ can be chosen in such a way that ${\cal F}_{1,2}$ yield
positive eigenvalues. However, in the presence of $V_{\slashed{K}}$, the minimum
of scalar potential of our model is shifted to Eq. (\ref{eq:mindev}).
The small deviations of
Eq. (\ref{eq:mindev}) can be computed from Eq. (\ref{eq:devexp}). From
Eq. (\ref{eq:devexp}), we can see that $\delta_-,\delta_+\neq0$. Hence,
$v_2\neq v_3$. Using the expressions for $\delta_-,\delta_+$ in Eq.
(\ref{eq:mutau}), we can get the required hierarchy between $m_\mu$ and
$m_\tau$, provided the parameters of $V_{\slashed{K}}$ are small. It can be
noticed that the parametrizations we have used in Eq. (\ref{eq:mindev})
is similar to that in Ref. \cite{gl2}. In Ref. \cite{gl2}, it is pointed
that $\delta_+=0$. In our work, we get $\delta_+\neq0$, since $f_+\neq0$.
This difference is due to the fact that in Ref. \cite{gl2}, $K$-violating
quartic terms are not considered.

We have described that with $K$-violating terms of our model, we can
explain
the required hierarchy between muon and tau lepton masses. However, in doing so,
from Eq. (\ref{eq:devexp}) we can see that $\delta_\theta\neq0$. This makes
$v_\Delta$ complex. One can fine tune the parameters in ${\cal F}_2,f_+,
f_\theta$ in such a way that $\delta_\theta=0$. On the other hand, to get
$\delta_\theta=0$, we can take ${\cal F}_{+\theta}=0=f_\theta$. After using
Eq. (\ref{eq:tcon}), ${\cal F}_{+\theta}=0$ implies
$\kappa_2=\kappa_3=0$. In order to make $f_\theta=0$, either we can take
$\delta\kappa_2=-\delta\kappa_2^\prime$ or forbid the trilinear  terms
of $V_{\slashed{K}}$. From the above made observations, we can see that,
in case I in order to make $v_\Delta$ real without
fine tuning the parameters, the trilinear
terms of $V_{total}$ containing $\phi_{2,3}$ should be forbidden.

\subsection{Case II}

As explained before, in this case, terms involving triplet Higgs give very
small contribution in comparison to that involving only doublet Higgses.
As a result of this, minimization of $V_{total}$ in this case proceeds
in two steps. In the first step, we minimize $V_{total}$ which contain only
doublet Higgses and thereby determine the VEVs of these fields. Later, after
using the VEVs of doublet Higgses, we minimize the potential containing the
triplet Higgs field. In the first step of minimization, we can neglect
$V_T$ in comparison to $V_D$ and also neglect the terms in $V_{\slashed{K}}$ which
contain triplet Higgs field. In this case also, we parametrize the VEVs for
$\phi_{1,2,3}$ and $\Delta$ as given by Eq. (\ref{eq:para}). Now, after
minimizing
$V_D$ with respect to $\sigma,\alpha,\beta$, the minimum is given
by Eq. (\ref{eq:min}) with the condition of Eq. (\ref{eq:dcon}). Since this
minimum gives $m_\mu=0$, we introduce $V_{\slashed{K}}$ and parametrize the
deviations in $\sigma,\alpha,\beta$ as given by Eq. (\ref{eq:mindev}).
After doing this, one can notice that the above mentioned deviations
can be found from Eq. (\ref{eq:dev}), where, in this case, ${\cal F}$ and $f$
are 3$\times$3 and 3$\times$1 matrices respectively. The components of
${\cal F}$ and $f$ can be found from Eqs. (\ref{eq:F}) and (\ref{eq:f}),
where one has to omit the terms containing $v^\prime$. As a result of this,
we get $\delta_-,\delta_+\neq0$. After using this in Eq. (\ref{eq:mutau}),
we get small and non-zero $m_\mu$.

Since the VEVs of doublet Higgses are determined, we now minimize $V_T$ and
try to see if $v_\Delta$ can be real. After using Eq. (\ref{eq:para}) in
$V_T$, we get
\begin{eqnarray}
\langle V_T\rangle&=&(m_\Delta^2+\lambda_{11}v_1^2+\lambda_{12}v^2)v^{\prime^2}
+\frac{1}{2}\lambda_\Delta v^{\prime^4}-2\kappa_1v_1^2v^\prime\cos\theta
\nonumber \\
&&-2\kappa_2v^2v^\prime[\cos^2\sigma\cos(\theta-2\alpha)
+\sin^2\sigma\cos(\theta-2\beta)]
+\kappa_3v^2v^\prime\sin2\sigma\sin(\theta-\alpha-\beta).
\nonumber \\
\end{eqnarray}
Since we are looking for a minimum at $\theta=0$, we do
\begin{equation}
\left.\frac{\partial\langle V_T\rangle}{\partial\theta}\right|_{\theta=0}=0
\Rightarrow -2\kappa_2[\cos^2\sigma\sin2\alpha+\sin^2\sigma\sin2\beta]
+\kappa_3\sin2\sigma\cos(\alpha+\beta)=0.
\end{equation}
As stated before, we have determined $\sigma,\alpha,\beta$ up to first
order in $\delta_0,\delta_-,\delta_+$. Plugging the parametrizations for
$\sigma,\alpha,\beta$ in the above equation and expanding the terms up to
first order in $\delta_0,\delta_-,\delta_+$, we get
\begin{equation}
2\kappa_2\sin2\omega-\kappa_3\cos2\omega
+2(2\kappa_2\cos2\omega+\kappa_3\sin2\omega)\delta_+=0.
\end{equation}
Since we have $\delta_+\neq0$, after equating the leading and subleading
terms of the above equation to zero, we get $\kappa_2=\kappa_3=0$. Hence,
in case II, in order to make $v_\Delta$ real, one has to forbid the
trilinear terms of $V_T$ which contain $\phi_{2,3}$.

\subsection{Imposing an extra $Z_3$ symmetry}

From the analysis of previous two subsections, we have seen that the
trilinear terms in $V_{total}$ which contain $\phi_{2,3}$ should be
forbidden in order to make $v_\Delta$ real. To achieve this, we impose
the discrete symmetry $Z_3$ in our model. Under this symmetry, the non-trivial
transformations are as follows.
\begin{eqnarray}
&&\phi_2\to\Omega\phi_2,\quad\phi_3\to\Omega\phi_3,
\nonumber \\
&&\mu_R\to\Omega^2\mu_R,\quad\tau_R\to\Omega^2\tau_R.
\end{eqnarray}
Here, $\Omega=e^{2\pi i/3}$. Under the above transformations, the Yukawa
couplings for leptons are invariant but the following couplings in $V_{total}$
are forbidden: $\lambda_{8,9},\kappa_{2,3},\delta\kappa,\delta\kappa^\prime$.
Now, after using the parametrizations of Eq. (\ref{eq:para}) in
$V_{inv}$, we get
\begin{equation}
\langle V_{inv}\rangle=\frac{1}{4}[\tilde{\lambda}-4\lambda_7\sin^2\zeta]v^4
\sin^22\sigma+\frac{1}{2}\lambda_{10}v^4\sin4\sigma\sin\zeta
-2\kappa_1v_1^2v^\prime\cos\theta.
\label{eq:z3vin}
\end{equation}
Here, $\zeta=\alpha-\beta$. In the above equation, we have neglected constant
terms which do not depend on $\sigma,\alpha,\beta,\theta$. We can notice from
the
above equation that $\theta$ do not mix with $\sigma,\zeta$. Moreover, due to
absence of trilinear terms in $V_{\slashed{K}}$, $\langle V_{\slashed{K}}\rangle$ do not
depend on $\theta$. As a result of this, we can see that $\theta=0$ is a
minimum to $V_{total}$ if $\kappa_1v^\prime>0$. This statement is true for
both the cases of I and II. Hence, after imposing the
above mentioned $Z_3$ symmetry, $v_\Delta$ can be real in our model.

For Eq. (\ref{eq:z3vin}), the minimum in terms of $\sigma,\zeta$ can be at
\begin{equation}
\sigma=\frac{\pi}{4},\quad\zeta=0.
\label{eq:z3vac}
\end{equation}
Since $\zeta=0$ corresponds to $m_\mu=0$, we introduce $K$-violating terms
into the model. As a result of this, the above mentioned minimum can be
shifted by small deviations $\delta_0,\delta_\zeta$ as
\begin{equation}
\sigma=\frac{\pi}{4}-\frac{\delta_0}{2},\quad\zeta=0+\delta_\zeta.
\end{equation}
Now, after imposing $Z_3$ symmetry in Eq. (\ref{eq:soft}) and after following
the procedure for minimizing $V_{total}$, which is described in Sec. 3.1,
we get
\begin{eqnarray}\label{eq:Z3-f0-fzeta}
\left(\begin{array}{c} \delta_0\\\delta_\zeta \end{array}\right)
&=&{\cal F}^{-1}\left(\begin{array}{c} f_0\\f_\zeta \end{array}\right),\quad
{\cal F}=\left(\begin{array}{cc}
-\frac{1}{2}\tilde{\lambda} & \lambda_{10}\\\lambda_{10} & -2\lambda_7
\end{array}\right)v^4,
\nonumber \\
f_0&=&\frac{1}{2}(\delta M_2^2-\delta M_3^2)v^2
+\frac{1}{2}(\delta\lambda_2-\delta\lambda_2^\prime)v^4
+\frac{1}{2}(\delta\lambda_3
-\delta\lambda_3^\prime+\delta\lambda_5-\delta\lambda_5^\prime)(v_1v)^2
\nonumber \\
&&+\frac{1}{2}(\delta\lambda_{12}-\delta\lambda_{12}^\prime)(vv^\prime)^2
\nonumber \\
f_\zeta&=&\delta M_s^2v^2
+\frac{1}{2}(\delta\lambda_{10}+\delta\lambda_{10}^\prime)v^4
+(\delta\lambda_s-\delta\lambda_s^\prime)(v_1v)^2
+\delta_t(vv^\prime)^2
\end{eqnarray}
We can see that $\delta_0,\delta_\zeta\neq0$. After using these in the
parametrizations for $v_{2,3}$, from Eq. (\ref{eq:mutau}), we get
\begin{equation}
\frac{m_\mu}{m_\tau}=\frac{1}{2}|\delta_0+i\delta_\zeta|.
\end{equation}
Using the above equation, the required hierarchy between muon and tau leptons
can be explained if we take $\delta_0,\delta_\zeta\sim0.1$.

In Sec. 2 we have described our model for lepton sector by introducing additional
fields and symmetries. In the current section, we have introduced one more symmetry,
$Z_3$, in order to make the triplet Higgs VEV to be real. In Tab.
\ref{table:lepton symmetries and roles} we
summarize the additional fields and symmetries, which are needed for our model, in the
lepton sector.
\begin{table}[h!]
	\begin{center}
		\begin{tabular}{|c|c|}
			\hline
			additional field &  role \\
			\hline
			$\phi_1$ & to generate the mass of electron \\
			\hline
			$\phi_2$, $\phi_3$ & to generate masses for $\mu$ and $\tau$\\
			\hline
			$\Delta$ & to generate masses for neutrinos\\
			\hline
		\end{tabular}
	\end{center}
	\begin{center}
		\begin{tabular}{|c|c|}
			\hline
			additional symmetry &  role \\
			\hline
			$CP$ symmetry & to get $\mu-\tau$ form for neutrino mass matrix \\
			\hline
			$Z_2$ & forbids unwanted Yukawa couplings among charged leptons\\
			\hline
			$U(1)_{L_\alpha}$ & to get diagonal masses for charged leptons\\
			\hline
			$K$ symmetry & to reduce the fine-tuning in muon and tau masses\\
			\hline
			$Z_3$ & to make the VEV of $\Delta$ to be real\\
			\hline
		\end{tabular}
	\end{center}
	\caption{Additional fields and symmetries, which are introduced in the lepton
	sector of our model. The roles of these fields and symmetries are also described
	here.}
	\label{table:lepton symmetries and roles}
\end{table}

\section{Neutrino mass ordering and the smallness of $\theta_{13}$}

After showing that the triplet Higgs can acquire real VEV, the neutrino
mass matrix of the model proposed in Sec. 2 satisfy Eq. (\ref{eq:mnu}). As a
result of this, after diagonalizing $M_\nu$, $\theta_{23}$ and
$\delta_{CP}$ would
be maximal \cite{gl}. However, the form of $M_\nu$ doesn't give predictions
about $\theta_{12},\theta_{13}$ and also about neutrino mass ordering. In this
section, we do an analysis and give a procedure which can give predictions about
neutrino mass ordering and the smallness of $\theta_{13}$ in our model.

In the model proposed in Sec. 2, the charged lepton masses are in diagonal
form. Hence,
the unitary matrix which diagonalizes $M_\nu$ can be written as
\begin{eqnarray}
&&U=\tilde{U}U_{PMNS},\quad \tilde{U}={\rm diag}(1,1,-1),
\nonumber \\
&&
U_{PMNS} = \left(\begin{array}{ccc}
c_{12}c_{13} & s_{12}c_{13} & s_{13}e^{-i\delta_{CP}} \\
-s_{12}c_{23}-c_{12}s_{23}s_{13}e^{i\delta_{CP}} &
c_{12}c_{23}-s_{12}s_{23}s_{13}e^{i\delta_{CP}} & s_{23}c_{13} \\
s_{12}s_{23}-c_{12}c_{23}s_{13}e^{i\delta_{CP}} &
-c_{12}s_{23}-s_{12}c_{23}s_{13}e^{i\delta_{CP}} & c_{23}c_{13}
\end{array}\right).
\label{eq:pmns}
\end{eqnarray}
Here, $c_{ij}=\cos\theta_{ij}$ and $s_{ij}=\sin\theta_{ij}$. $U_{PMNS}$ is
the Pontecorvo-Maki-Nakagawa-Sakata (PMNS) matrix which is parameterized
in terms of the three
lepton mixing angles and the $CP$ violating Dirac phase, according to the
convention of PDG \cite{pdg}. Diagonal elements of $\tilde{U}$ can be absorbed
into the charged lepton fields. Now, the relation for diagonalizing $M_\nu$
can be expressed as
\begin{equation}
M_\nu=U^*{\rm diag}(m_1,m_2,m_3)U^\dagger.
\label{eq:mdia}
\end{equation}
While solving the above equation, we can use an approximation procedure
\cite{prwo} which
is related to neutrino masses and the mixing angle $\theta_{13}$. This
procedure is explained below.

In the expression for $U_{PMNS}$ one can have Majorana phases. These phases cannot be
determined from neutrino oscillation data. But they can affect the life-time of
neutrinoless double beta decay, since neutrinos in our model are Majorana particles.
However, so far no concrete evidence is there for this decay \cite{pdg}, and as a
result, the Majorana phases can be any where between 0 and $2\pi$. Hence, in our
analysis, for the sake of simplicity, we have chosen these phases to be zero. On
the other hand, by taking some specific values for Majorana phases in the below
described procedure, one can study the conditions which can give rise for neutrino
Yukawa couplings of our model. However, we reserve this study for future.

In $U_{PMNS}$, we put $\theta_{23}=\frac{\pi}{4}$ and
$\delta_{CP}=\frac{3\pi}{2}$.
From the neutrino oscillation data, we have $s_{12}^2\sim\frac{1}{3}$ and
$s_{13}^2\sim2\cdot10^{-2}$ \cite{glo-fit}. Here we can notice that $s_{13}^2$
is negligibly small in comparison to unity, and hence $s_{13}\sim0.15$ can
be treated
as small variable. On the other hand, $s_{12}^2$ and $s_{23}^2$ are
of order one. Since $s_{13}$ is the only small variable in $U_{PMNS}$,
we expand $U_{PMNS}$ up to first order in $s_{13}$. Expression for this
is given below.
\begin{eqnarray}\label{eq:upmns expansion}
&&U_{PMNS}=U_0+\delta U,
\nonumber \\
&&U_0=\left(\begin{array}{ccc}
c_{12} & s_{12} & 0 \\
-\frac{s_{12}}{\sqrt{2}} & \frac{c_{12}}{\sqrt{2}} & \frac{1}{\sqrt{2}} \\
\frac{s_{12}}{\sqrt{2}} & -\frac{c_{12}}{\sqrt{2}} & \frac{1}{\sqrt{2}}
\end{array}\right),\quad
\delta U=\left(\begin{array}{ccc}
0 & 0 & 1 \\
\frac{c_{12}}{\sqrt{2}} & \frac{s_{12}}{\sqrt{2}} & 0 \\
\frac{c_{12}}{\sqrt{2}} & \frac{s_{12}}{\sqrt{2}} & 0
\end{array}\right)is_{13}.
\end{eqnarray}
From the neutrino oscillation data, two mass-squared differences for neutrinos
are found, which are given in Eq. (\ref{eq:msquare}). From this equation
we can notice that
$\frac{m_s^2}{m_a^2}\sim s_{13}^2$, which is negligibly small in comparison
to unity. This indicates, an approximation with
respect to neutrino masses can also be applied while solving the Eq.
(\ref{eq:mdia}). In order to fit the mass-square differences of
Eq. (\ref{eq:msquare}), we can take the neutrino masses as follows.
\begin{eqnarray}
&& {\rm NO}:\quad m_1\lapprox m_s,\quad m_2=\sqrt{m_s^2+m_1^2},\quad
m_3=\sqrt{m_a^2+m_1^2}.
\nonumber \\
&& {\rm IO}:\quad m_3\lapprox m_s,\quad m_1=\sqrt{m_a^2+m_3^2},\quad
m_2=\sqrt{m_s^2+m_1^2}.
\label{eq:numval}
\end{eqnarray}
Now we can notice that $\frac{m_1}{m_a}\lapprox s_{13}$ in the case of NO.
Whereas, $\frac{m_1}{m_a}\sim1$ in the case of IO. Similar
conclusions can be made about $\frac{m_2}{m_a}$ and $\frac{m_3}{m_a}$.

Using the approximation scheme described in the previous paragraph, we
can expand $\frac{1}{m_a}M_\nu$ in powers of $s_{13},\frac{m_s}{m_a}$.
After neglecting the second and higher order corrections in
$s_{13},\frac{m_s}{m_a}$, for both the cases of NO and IO,
the elements of $\frac{1}{m_a}M_\nu$ are given below.
\begin{eqnarray}
&&\frac{1}{m_a}M_\nu=\frac{1}{2m_a}\left(\begin{array}{ccc}
x & z & z^*\\
z & w & y\\
z^* & y & w^*
\end{array}\right),
\nonumber \\
{\rm NO}&:&x=2c_{12}^2m_1+2 s_{12}^2 m_2,\quad
z=\sqrt{2} c_{12}s_{12}(m_2-m_1)-i\sqrt{2}m_3s_{13},
\nonumber \\
&&w=m_3+c_{12}^2m_2+s_{12}^2m_1,\quad
y=-m_3+c_{12}^2m_2+s_{12}^2m_1.
\nonumber \\
{\rm IO}&:&x=2m_1,\quad
z=-\sqrt{2} i s_{13}m_1,\quad w=m_1+m_3,\quad
y=m_1-m_3.
\label{cond}
\end{eqnarray}
Using the above relations, in order for the matrix $M_\nu$ to predict about
neutrino mass ordering and smallness of $\theta_{13}$, the Yukawa couplings
in Eq. (\ref{eq:lagnu}) should satisfy the following conditions.
\begin{itemize}
\item To predict NO and smallness of $\theta_{13}$:
\begin{enumerate}[(i)]
\item $Y^\nu_{ee},Y^\nu_{e\mu}$ should be suppressed by about 0.1
as compared to that of $Y^\nu_{\mu\mu},Y^\nu_{\mu\tau}$.
\item $Y^\nu_{\mu\mu}$ should be real.
\end{enumerate}
\item To predict IO and smallness of $\theta_{13}$:
\begin{enumerate}[(i)]
\item $Y^\nu_{e\mu}$ should be purely imaginary and its magnitude is suppressed
by about 0.1 as compared to other elements of $Y^\nu$.
\item $Y^\nu_{\mu\mu}$ should be real.
\end{enumerate}
\end{itemize}
It is to remind here that $Y^\nu$ is a symmetric matrix and satisfies
Eq. (\ref{eq:ynu}). Hence, not all elements of $Y^\nu$ are independent.
As a result of this, while describing the above conditions, we have considered
$Y^\nu_{ee},Y^\nu_{e\mu},Y^\nu_{\mu\mu},Y^\nu_{\mu\tau}$ as independent
elements of $Y^\nu$. Another point to mention here is that the above
mentioned conditions are true after neglecting second and higher order
corrections in $\frac{1}{m_a}M_\nu$.

The condition (ii) described for the cases of NO and IO is trivially satisfied if one
uses relations of Eq. (\ref{cond}). The non-trivial condition to check is the
condition (i) in the
cases of both NO and IO. The suppression factor mentioned in this condition is arising due
to $s_{13}\sim\frac{m_s}{m_a}\sim 0.15$. We have checked this suppression factor for
the case of NO by computing the following ratios:
$\frac{|Y^\nu_{ee}|}{|Y^\nu_{\mu\mu}|}$, $\frac{|Y^\nu_{ee}|}{|Y^\nu_{\mu\tau}|}$,
$\frac{|Y^\nu_{e\mu}|}{|Y^\nu_{\mu\mu}|}$, $\frac{|Y^\nu_{e\mu}|}{|Y^\nu_{\mu\tau}|}$.
While for the case of IO, the following ratios are computed in order to check
condition (i): $\frac{|Y^\nu_{e\mu}|}{|Y^\nu_{ee}|}$,
$\frac{|Y^\nu_{e\mu}|}{|Y^\nu_{\mu\mu}|}$, $\frac{|Y^\nu_{e\mu}|}{|Y^\nu_{\mu\tau}|}$.
One can notice that the neutrino Yukawa couplings are proportional to the elements of
$M_\nu$, which are given in Eq. (\ref{cond}). The neutrino masses in Eq. (\ref{cond})
are computed by using Eq. (\ref{eq:numval}) and also by varying $m^2_s,m^2_a$
over their allowed $3\sigma$ ranges. The mass of lightest neutrino is varied from  0 to
$m_s$ in both the cases of NO and IO. While computing the above mentioned ratios,
we have also varied $s_{12}^2$
and $s_{13}^2$ over their allowed $3\sigma$ ranges. We have tabulated the allowed $3\sigma$
ranges for the above mentioned variables in Tab. \ref{table:allowed ranges}.
\begin{table}[h!]
	\begin{center}
		\begin{tabular}{c |c}
			\hline\hline
			parameters & allowed range \\
			\hline\hline
			$m_s^2$& (6.94$-$8.14)$\times 10^{-5}$ eV$^2$ \\
			\hline
			$m_a^2$ (NO) & (2.47$-$2.63)$\times 10^{-3}$ eV$^2$\\
			\hline
			$m_a^2$ (IO)& (2.37$-$2.53)$\times 10^{-3}$ eV$^2$\\
			\hline
			$s_{12}^2$ & 0.271$-$0.369 \\
			\hline
			$s_{13}^2$ (NO) & 0.0200$-$0.02405 \\
			\hline
			$s_{13}^2$ (IO) & 0.02018$-$0.02424 \\
			\hline\hline
		\end{tabular}
	\end{center}
	\caption{Allowed $3\sigma$ ranges of the neutrino oscillation observables \cite{glo-fit}, which are used in our analysis.}
	\label{table:allowed ranges}
\end{table}

While doing
the above described analysis, we have also checked if the sum of the three neutrinos
is less than 0.12 eV, which is a constraint obtained from the cosmological observations
\cite{cosmo}. As already described in the previous paragraph, the suppression in the
ratios of various Yukawa couplings should be around $s_{13}\sim\frac{m_s}{m_a}\sim 0.15$.
However, in the analysis we have found that some of these ratios can become as large
as 0.5, and thus, invalidate the approximation procedure that we are using here.
Hence, in the analysis we have restricted all these ratios
to be less than or of the order of 0.2. Selected plots from this analysis are
presented in Fig. 1.
\begin{figure}[h!]
	\begin{center}
		\includegraphics[height=55mm,width=75mm]{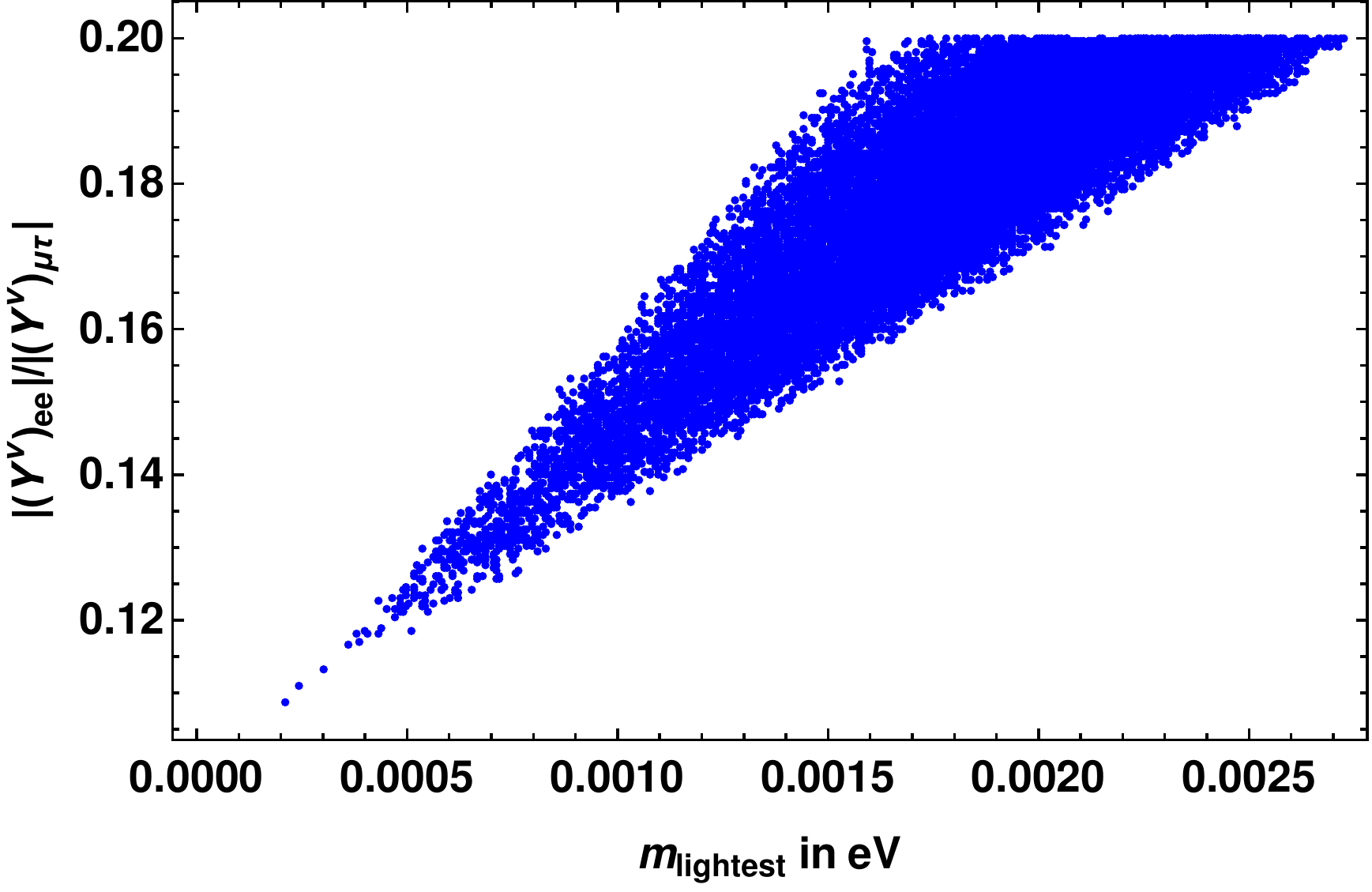}
		\includegraphics[height=55mm,width=75mm]{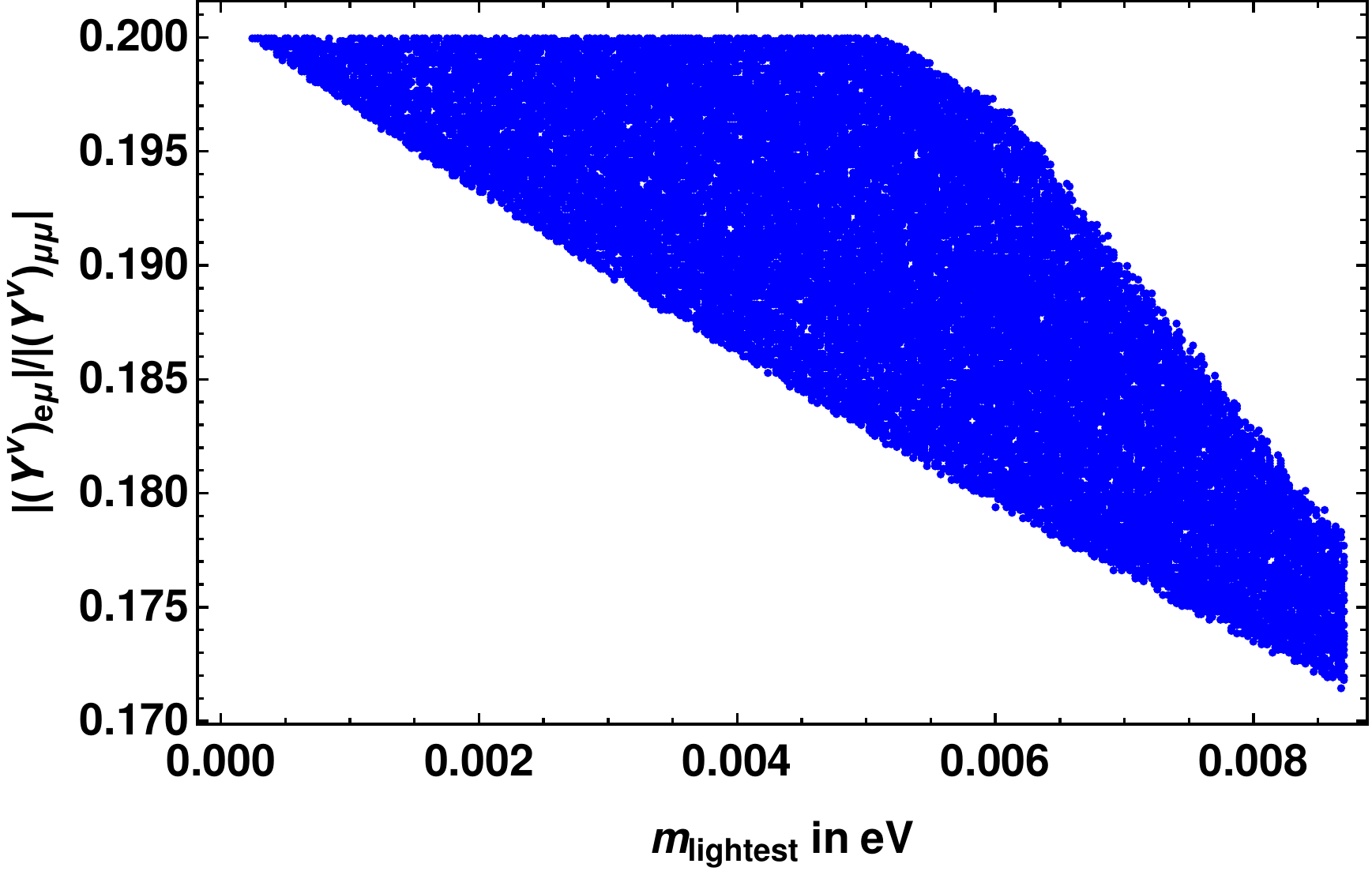}
	\end{center}
	\caption{Ratios of the magnitude of Yukawa couplings versus the mass of lightest
	neutrino. The left- and right-hand side plots are for NO and IO respectively.
	In both the plots we have varied the neutrino oscillation observables over their
	allowed $3\sigma$ ranges, which are given in Tab. \ref{table:allowed ranges}.
	For other details, see the text.}
	\label{fig:Yukawa ratios}
\end{figure}
From this figure we can see that the mass of lightest neutrino, $m_{\rm lightest}$,
is constrained due to above mentioned restriction on the ratios of Yukawa couplings.
We can notice that $m_{\rm lightest}$ has narrow allowed region in the case of NO as
compared to that of IO. Apart from the $m_{\rm lightest}$ in Fig. 1, $s_{13}^2$ is also
constrained to be in the range of 0.02 to 0.023, in the case of NO. But otherwise, this
variable is not constrained in the case of IO. As for $s_{12}^2$, we have found that
it can take the full $3\sigma$ range in both the cases of NO and IO of Fig. 1. Although
we have presented selected plots in Fig. 1, we have got similar kind of plots for other
ratios of Yukawa couplings, which are mentioned in the previous paragraph. These plots
justify the approximation procedure that we are using here and also
verify the condition (i) for the cases of NO and IO, which are mentioned below
Eq. (\ref{cond}).

The conditions mentioned below Eq. (\ref{cond}), for both the cases of NO and IO,
cannot be achieved just
with the $CP$ symmetry. Additional mechanism should be proposed in order to
satisfy these conditions. In an attempt towards this, we have given one
mechanism to achieve condition (i) for the case of NO, where the necessary
suppression in the Yukawa couplings is explained through non-renormalizable
terms within the framework of $CP$ symmetry. This mechanism is presented
in the Appendix.
In this work, we do not have a mechanism to achieve condition (i) for the case
of IO and also to achieve condition (ii) for both NO and IO. We work on these
problems in future. As already described, with generalized $CP$ transformations
and $\mu-\tau$ symmetry, $\theta_{23}$ and $\delta_{CP}$ will be maximal.
However, in future neutrino oscillation experiments,
$\theta_{23}$ and $\delta_{CP}$ may be found to be away from their maximal
values. In such a case, one needs to device a mechanism for the breaking of
$\mu-\tau$ reflection symmetry in order to explain the non-maximal values
for the above mentioned observables. These topics are outside the scope of
this paper.

\section{Quark mixing}

In the model we proposed for lepton mixing, three Higgs doublets exist. Since
these Higgs doublets can also give masses to quarks, it is interesting to see if
quark mixing can also be explained with $CP$ and other symmetries
of our model. As already described in Sec. 1, since there is a hierarchy among
quark masses, the mixing pattern for quarks can be explained if their
Yukawa couplings are hierarchically suppressed. Babu and Nandi have proposed
one model \cite{bn} for explaining quark mixing through hierarchically
suppressed Yukawa couplings. Later, this model has been modified in Ref.
\cite{lmn}, where the suppression in Yukawa couplings is explained with a
singlet scalar field. We follow the work of Refs. \cite{bn,lmn} in order to
explain quark mixing in our framework.

\subsection{Model for quark masses and mixing}

We denote the three families of quark doublets, up- and down-type singlets
as $Q_{jL}$, $u_{jR}$ and $d_{jR}$, respectively. We propose a scalar field
$X$ which is singlet under standard model gauge group. \textbf{We assume all the quark
doublets to be singlets under the symmetry $K\times Z_2\times Z_3$. On the other hand, all the right-handed quark fields are singlets under $K\times Z_3$ but they are odd under $Z_2$ symmetry}. $X$ field
is singlet under $K\times Z_3$ but is odd under $Z_2$.
Both the quark and $X$ fields transform under $CP$ symmetry as
\begin{equation}
Q_{jL}\to i\gamma^0C\bar{Q}_{jL}^T,\quad
u_{jR}\to i\gamma^0C\bar{u}_{jR}^T, \quad d_{jR}\to i\gamma^0C\bar{d}_{jR}^T,
\quad X\to X^*.
\label{eq:CP transf quarks}
\end{equation}
Now, with the above mentioned transformations and fields, we
consider the following effective Lagrangian for quark masses.
\begin{eqnarray}
{\cal L}_Y&=&h^u_{33}\bar{Q}_{3L}\tilde{\phi}_1u_{3R}
+\left(\frac{X}{M}\right)^2[h^d_{33}\bar{Q}_{3L}\phi_1 d_{3R}
+h^u_{22}\bar{Q}_{2L}\tilde{\phi}_1u_{2R}
+h^u_{23}\bar{Q}_{2L}\tilde{\phi}_1u_{3R}
\nonumber \\
&& +h^u_{32}\bar{Q}_{3L}\tilde{\phi}_1 u_{2R}]+\left(\frac{X}{M}\right)^4[h^d_{22}\bar{Q}_{2L}\phi_1 d_{2R}
+h^d_{23}\bar{Q}_{2L}\phi_1 d_{3R}+h^d_{32}\bar{Q}_{3L}\phi_1 d_{2R}
\nonumber \\
&&+h^u_{12}\bar{Q}_{1L}\tilde{\phi}_1u_{2R}+h^u_{21}\bar{Q}_{2L}\tilde{\phi}_1u_{1R}+h^u_{13}\bar{Q}_{1L}\tilde{\phi}_1u_{3R}+ h^u_{31}\bar{Q}_{3L}\tilde{\phi}_1u_{1R}]\nonumber \\
&&+\left(\frac{X}{M}\right)^6[h^u_{11}\bar{Q}_{1L}\tilde{\phi}_1u_{1R} +h^d_{11}\bar{Q}_{1L}\phi_1 d_{1R}+h^d_{12}\bar{Q}_{1L}\phi_1 d_{2R} +h^d_{13}\bar{Q}_{1L}\phi_1 d_{3R}]
\nonumber \\
&&+\left(\frac{X}{M}\right)^{10}[h^d_{21}\bar{Q}_{2L}\phi_1 u_{1R}]
+\left(\frac{X}{M}\right)^{10}[h^d_{31}\bar{Q}_{3L}\phi_1 u_{1R}]+h.c.
\label{eq:qlag}
\end{eqnarray}
The above Lagrangian is valid below a mass scale of $M$. The
non-renormalizable terms of this Lagrangian can be motivated from
the UV completion of this model, which is presented in the next subsection.
According to this UV completion, we propose a flavor symmetry $U(1)_F$ and
heavy vector-like quark (VLQ) fields above the scale $M$. After integrating
the heavy fields of our model, below the scale $M$, the non-renormalizable
terms of Eq. (\ref{eq:qlag}) can appear. Here we can see that $M$ represents
the mass scale of heavy VLQs. Since new particles can be probed at LHC
experiment if their masses are around 1 TeV, so we take $M\sim$ 1 TeV.

Due to $CP$ symmetry, the Yukawa
couplings $h^{u,d}_{jk}$ should be real in Eq. (\ref{eq:qlag}). After
$X$ acquires VEV, for $\langle X\rangle<M$, we can see that
$\frac{X}{M}$ gives suppression to effective quark Yukawa couplings.
Since the Yukawa couplings of Eq. (\ref{eq:qlag}) are real, we assume
$\langle X\rangle$ is complex and this can be the source for $CP$ violation
in the quark sector. In the Lagrangian of Eq. (\ref{eq:qlag}), only the
doublet $\phi_1$ generates Yukawa couplings for quark fields. The other
doublets $\phi_{2,3}$ do not generate these Yukawa couplings due to
the presence of $CP\times K$ symmetry.

After electroweak symmetry breaking, using Eq. (\ref{eq:qlag}),
the matrices for up- and down-type quarks can be written, respectively, as
\begin{equation}
M_u=\left(\begin{array}{ccc}
h^u_{11}\epsilon^6 & h^u_{12}\epsilon^4 & h^u_{13}\epsilon^4 \\
h^u_{21}\epsilon^4 & h^u_{22}\epsilon^2 & h^u_{23}\epsilon^2 \\
h^u_{31}\epsilon^4 & h^u_{32}\epsilon^2 & h^u_{33}
\end{array}\right)v_1,\quad
M_d=\left(\begin{array}{ccc}
h^d_{11}\epsilon^6 & h^d_{12}\epsilon^6 & h^d_{13}\epsilon^6 \\
h^d_{21}\epsilon^{10} & h^d_{22}\epsilon^4 & h^d_{23}\epsilon^4 \\
h^d_{31}\epsilon^{10} & h^d_{32}\epsilon^4 & h^d_{33}\epsilon^2
\end{array}\right)v_1.
\label{eq:qmas}
\end{equation}
Here, $\epsilon=\frac{\langle X\rangle}{M}$. The form of $M_{u,d}$ is similar
to the corresponding matrices of Refs. \cite{bn,lmn}. However, the only
difference is that the elements 21 and 31 of $M_d$ are generated at higher
order as compared to that of Refs. \cite{bn,lmn}. It is argued
in Ref. \cite{lmn} that the above mentioned elements do not affect quark
masses and mixing if they are generated at higher order.
Hence, after diagonalizing the above matrices, the masses and
mixing angles for quarks, up to leading order in $|\epsilon|$, are given by
\begin{eqnarray}
&&(m_t,m_c,m_u)\approx(|h^u_{33}|,|h^u_{22}||\epsilon|^2,
|h^u_{11}-h^u_{12}h^u_{21}/h^u_{22}||\epsilon|^6)v_1,
\nonumber \\
&&(m_b,m_s,m_d)\approx(|h^d_{33}||\epsilon|^2,
|h^d_{22}||\epsilon|^4,|h^d_{11}||\epsilon|^6)v_1,
\nonumber \\
&&|V_{us}|\approx\left|\frac{h^d_{12}}{h^d_{22}}-
\frac{h^u_{12}}{h^u_{22}}\right||\epsilon|^2,
\nonumber \\
&&|V_{cb}|\approx\left|\frac{h^d_{23}}{h^d_{33}}-
\frac{h^u_{23}}{h^u_{33}}\right||\epsilon|^2,
\nonumber \\
&&|V_{ub}|\approx\left|\frac{h^d_{13}}{h^d_{33}}-
\frac{h^u_{12}h^d_{23}}{h^u_{22}h^d_{33}}-
\frac{h^u_{13}}{h^u_{33}}\right||\epsilon|^4,
\nonumber \\
&&{\rm arg}(V_{ub})\approx4{\rm arg}(\epsilon).
\label{eq:qrel}
\end{eqnarray}

Due to three Higgs doublets in our model, we have $|v_1|^2+|v_2|^2+|v_3|^2
\approx(174~{\rm GeV})^2$. To satisfy this, we take $v_1,v\sim174/\sqrt{2}$
GeV. With this value for $v_1$, we have fitted the expressions of
Eq. (\ref{eq:qrel}) to the following best fit values \cite{pdg}.
\begin{eqnarray}
&&(m_t,m_c,m_u)=(172.76,1.27,2.16\times10^{-3})~{\rm GeV},
\nonumber \\
&&(m_b,m_s,m_d)=(4.18\times10^3,93,4.67)~{\rm MeV},
\nonumber \\
&&(|V_{us}|,|V_{cb}|,|V_{ub}|)=(0.2245,0.041,0.00382),
\nonumber \\
&&{\rm arg}(V_{ub})=-1.196
\end{eqnarray}
After doing the above mentioned fitting, below we have given a sample
set of numerical values with $|\epsilon|=1/5.5$.
\begin{eqnarray}
&&(|h^u_{33}|,|h^u_{22}|,
|h^u_{11}-h^u_{12}h^u_{21}/h^u_{22}|)
\approx(1.4,0.31,0.49),
\nonumber \\
&&(|h^d_{33}|,|h^d_{22}|,|h^d_{11}|)\approx
(1.03,0.69,1.05),
\nonumber \\
&&(h^d_{12},h^u_{12},h^d_{23},h^u_{23},
h^d_{13},h^u_{13})\approx(1.49,-1.45,0.69,-0.8,1.12,1.0),
\nonumber \\
&&{\rm arg}(\epsilon)\approx-0.3
\label{eq:numer}
\end{eqnarray}
From the numerical values given above, we can see that the magnitudes of
all Yukawa couplings are less than about 1.5.
We have tried the numerical values with $|\epsilon|=1/6$. However, in 
this case some of the Yukawa couplings can become larger than 2.0.
Hence, with $|\epsilon|=1/5.5$ and ${\cal O}(1)$ Yukawa couplings,
we can explain the quark masses and mixing pattern in our model.
Since we expect new physics to appear around 1 TeV,
we can take the cut-off scale of Eq.(\ref{eq:qlag}) to be $M\sim$ 1 TeV.
Now, for $|\epsilon|=1/5.5$, we get $|\langle X\rangle|\sim$ 181 GeV.

\subsection{UV completion}

Here we present the UV completion for our model in order to explain
the origin of non-renormalizable terms of Eq. (\ref{eq:qlag}). To
achieve this UV completion, we follow the works of Refs. \cite{lmn,gh}.
The idea of this UV completion is to explain non-renormalizable terms following
from a theory which is renormalizable at a high scale. Hence, we assume our
model is renormalizable at and above the scale $M$ and propose a flavor
symmetry $U(1)_F$ which is exact above $M$. To generate non-renormalizable
terms below $M$, we propose additional fields like flavons and VLQs,
which transform under $U(1)_F$. The standard model quarks are charged
under the $U(1)_F$ symmetry. But the Higgs
doublets and $X$ field are singlets under $U(1)_F$. The $U(1)_F$ is
spontaneously broken when the flavons acquire VEVs around $M$, which is also
the mass scale of VLQs. Here, we can see that our model should
respect the symmetry $CP\times K\times Z_2\times Z_3\times U(1)_F$ above
the scale $M$. However, below $M$, after integrating the heavy
VLQs and flavon fields, our model should generate
non-renormalizable terms of Eq.(\ref{eq:qlag}), which respect the symmetry
$CP\times K\times Z_2\times Z_3$.

Under the $U(1)_F$, we denote the charges for  $Q_{jL}$, $u_{jR}$ and
$d_{jR}$ as $q_{jf}$, $u_{jf}$ and $d_{jf}$, respectively. We propose
only two flavon fields, $F_1$ and $F_2$, whose charges under $U(1)_F$ are
$f_1$ and $f_2$, respectively. Flavons are charged under $CP$ symmetry, but
otherwise are singlets under $K\times Z_2\times Z_3$.
Under the $CP$ symmetry, flavons transform like the $X$ field of our model.
Now, to generate non-renormalizable terms for
up-type quarks of Eq. (\ref{eq:qlag}), we introduce VLQs
$K_{jL}$ and $K_{jR}$, which are color triplets and their hypercharges
are same as that of right-handed singlet up-quarks. Analogous to
$K_{jL}$ and $K_{jR}$, we introduce $G_{jL}$ and $G_{jR}$, which
generate non-renormalizable terms for down-type quarks. The above VLQs
are singlets under $SU(2)$ symmetry of standard model and $K\times Z_3$.
These fields are charged under $Z_2$ symmetry. Under the $CP$ symmetry, they
transform like the quark fields.

After describing the field content and their charge assignments in the
UV completion of our model, below we explain the generation of
non-renormalizable terms of Eq. (\ref{eq:qlag}). The $h^u_{33}$ term of
Eq. (\ref{eq:qlag}) is renormalizable, which can be generated in our model
by taking $q_{3f}=u_{3f}$.
To generate $h^u_{32}$ term of Eq. (\ref{eq:qlag}), we consider the below
invariant terms in the UV completion of our model.
\begin{eqnarray}
\mathcal{L}^u_{32}&=&	\bar{Q}_{3L}\tilde{\phi}_1 K_{1R}+  F_1^*\bar{K}_{1R}K_{1L}+ X \bar{K}_{1L}K_{2R}+  F_2 \bar{K}_{2R}K_{2L}+  X \bar{K}_{2L}u_{2R}+h.c..
\label{eq:h32u-UV-Lag}
\end{eqnarray}
Since the terms in the above equation are invariant under $CP$ symmetry,
the dimensionless Yukawa couplings should be real. These Yukawa couplings
are ${\cal O}(1)$, which we have not written
explicitly here. The $U(1)_F$ charges for $K_{jL},K_{jR}$ can be fixed
in terms of corresponding charges of quarks and flavons in such a way
that the above equation is invariant under $U(1)_F$. Similarly, the $Z_2$
charges for these VLQs can be assigned so that the above equation is invariant
under $Z_2$. The $U(1)_F\times Z_2$ charges for VLQs of $K$-type are given in
Eq. (\ref{eq:U(1)F charges}). Now, when the flavons acquire VEVs, the VLQs
in Eq. (\ref{eq:h32u-UV-Lag}) acquire masses of the order of $M$. After
integrating these heavy VLQs, terms in Eq. (\ref{eq:h32u-UV-Lag}) generate
the $h^u_{32}$ term of Eq. (\ref{eq:qlag}).

By introducing more VLQs of $K$-type, the process described in the
previous paragraph can be applied in order to generate
other non-renormalizable terms of Eq. (\ref{eq:qlag}). Below we have
given the invariant Lagrangians of the form ${\cal L}^u_{ij}$, which generate
the $h^u_{ij}$ term of Eq. (\ref{eq:qlag}), after integrating the heavy
VLQs and flavons. The $U(1)_F\times Z_2$ charges for the VLQs in these
Lagrangians can be seen in Eq. (\ref{eq:U(1)F charges}).
\begin{eqnarray}\label{eq:h31u-UV-Lag}
\mathcal{L}^u_{31}&=&	\bar{Q}_{3L}\tilde{\phi}_1 K_{1R}+ F_1^*\bar{K}_{1R}K_{1L}+ X \bar{K}_{1L}K_{2R}+  F_2 \bar{K}_{2R}K_{2L}+ X \bar{K}_{2L}K_{3R}+ F_2 \bar{K}_{3R}K_{3L}\nonumber \\
	&&	+ X \bar{K}_{3L}K_{4R}+ M \bar{K}_{4R}K_{4L}+X\bar{K}_{4L}u_{1R}+h.c..
\end{eqnarray}
\begin{eqnarray}\label{eq:h23u-UV-Lag}
\mathcal{L}^u_{23}&=& \bar{Q}_{2L}\tilde{\phi}_1K_{5R}+M\bar{K}_{5R}K_{5L}+X \bar{K}_{5L}K_{6R}+ F_1 \bar{K}_{6R}K_{6L}+X \bar{K}_{6L}u_{3R}+h.c.
\end{eqnarray}
\begin{eqnarray}\label{eq:h22u-UV-Lag}
\mathcal{L}^u_{22}&=&	\bar{Q}_{2L}\tilde{\phi}_1K_{5R}+M\bar{K}_{5R}K_{5L}+ X\bar{K}_{5L}K_{7R}+F_2 \bar{K}_{7R}K_{7L}+X\bar{K}_{7L}u_{2R}+h.c.
\end{eqnarray}
\begin{eqnarray}\label{eq:h21u-UV-Lag}
\mathcal{L}^u_{21}&=&\bar{Q}_{2L}\tilde{\phi}_1K_{5R}+M\bar{K}_{5R}K_{5L}+X\bar{K}_{5L}K_{7R}+F_2 \bar{K}_{7R}K_{7L}+X \bar{K}_{7L}K_{8R}+M\bar{K}_{8R}K_{8L}+ \nonumber \\
&&	+X\bar{K}_{8L}K_{9R}+F_2\bar{K}_{9R}K_{9L}+X\bar{K}_{9L}u_{1R}+h.c.
\end{eqnarray}
\begin{eqnarray}\label{eq:h11u-UV-Lag}
\mathcal{L}^u_{11}&=& \bar{Q}_{1L}\tilde{\phi}_1K_{10R}+F_2 \bar{K}_{10R}K_{10L}+X\bar{K}_{10L}K_{11R}+F_2 \bar{K}_{11R}K_{11L} +X\bar{K}_{11L}K_{12R}\nonumber \\
&&	F_2\bar{K}_{12R}K_{12L}+X\bar{K}_{12L}K_{13R}+F_2 \bar{K}_{13R}K_{13L}+X\bar{K}_{13L}K_{14R}+F_2 \bar{K}_{14R}K_{14L}\nonumber \\
&& + X\bar{K}_{14L}K_{15R}+M\bar{K}_{15R}K_{15L}+X\bar{K}_{15L}u_{1R}+h.c..
\end{eqnarray}
\begin{eqnarray}\label{eq:h12u-UV-Lag}
\mathcal{L}^u_{12}&=& \bar{Q}_{1L}\tilde{\phi}_1K_{10R}+F_2 \bar{K}_{10R}K_{10L}+X\bar{K}_{10L}K_{11R}+F_2 \bar{K}_{11R}K_{11L}+X\bar{K}_{11L}K_{12R}\nonumber \\
	&&+F_2\bar{K}_{12R}K_{12L}+X\bar{K}_{12L}K_{13R}+F_2 \bar{K}_{13R}K_{13L}+X\bar{K}_{13L}u_{2R}+h.c..
\end{eqnarray} 
\begin{eqnarray}\label{eq:h13u-UV-Lag}
\mathcal{L}^u_{13}&=&	\bar{Q}_{1L}\tilde{\phi}_1K_{10R}+F_2 \bar{K}_{10R}K_{10L}+X\bar{K}_{10L}K_{11R}+F_2 \bar{K}_{11R}K_{11L}+X\bar{K}_{11L}K_{12R}\nonumber \\
&&+F_2\bar{K}_{12R}K_{12L}+X\bar{K}_{12L}K_{16R}+F_1 \bar{K}_{16R}K_{16L}+X\bar{K}_{16L}u_{3R}+h.c..
\end{eqnarray}
\begin{eqnarray}
U(1)_F&:& K_{1R}\rightarrow q_{3f},\quad K_{1L}, K_{2R}\rightarrow q_{3f}+f_1,\quad K_{2L}, K_{3R}\rightarrow q_{3f}+f_1-f_2, \nonumber \\
&& K_{3L},K_{4L}, K_{4R}\rightarrow q_{3f}+f_1-2f_2,\quad K_{5R}, K_{5L},K_{6R},K_{7R}\rightarrow q_{2f}\nonumber \\
&&  K_{7L},K_{8R},K_{8L},K_{9R}\rightarrow q_{2f}-f_2,\quad K_{6L}\rightarrow q_{2f}-f_1,\quad K_{9L}\rightarrow q_{2f}-2 f_2,\nonumber \\
&& K_{10R}\rightarrow q_{1f},\quad K_{10L},K_{11R}\rightarrow q_{1f}-f_2,\quad K_{11L},K_{12R}\rightarrow q_{1f}-2 f_2,\nonumber \\	
&& K_{12L},K_{13R},K_{16R}\rightarrow q_{1f}-3 f_2,\quad K_{13L},K_{14R}\rightarrow q_{1f}-4 f_2,\nonumber \\
&& K_{14L},K_{15R},K_{15L}\rightarrow q_{1f}-5 f_2,\quad K_{16L},u_{3R}\rightarrow q_{1f}-3 f_2-f_1.\nonumber \\
Z_2 &:& K_{1L}, K_{1R}, K_{3L}, K_{3R},K_{5L}, K_{5R},K_{8L},K_{8R}, K_{10L},K_{10R},K_{12L},K_{12R},K_{14L},K_{14R},\nonumber \\
&&\rightarrow \text{odd},\quad K_{2L}, K_{2R}, K_{4L}, K_{4R},K_{6L},K_{6R},K_{7L}, K_{7R},K_{9L},K_{9R},K_{11L},K_{11R}\nonumber \\
&&,K_{13L},K_{13R},K_{15R},K_{15L},K_{16L},K_{16R}\rightarrow \text{even}.
\label{eq:U(1)F charges}
\end{eqnarray} 

Since the Lagragians of Eqs. (\ref{eq:h32u-UV-Lag}) -
(\ref{eq:h13u-UV-Lag}) are invariant under $U(1)_F$, we get relations among
the $U(1)_F$ charges of quarks and flavons. These relations can be consistently
solved. Taking $q_{3f}$, $f_1$ and $f_2$ as independent variables, the above
mentioned relations can be expressed as
\begin{eqnarray}
&&q_{2f}=q_{3f}+f_1,\quad q_{1f}=q_{3f}+f_1+3f_2,
\nonumber \\
&&u_{3f}=q_{3f},\quad u_{2f}=q_{3f}+f_1-f_2,\quad u_{1f}=q_{3f}+f_1-2f_2.
\label{eq:urel}
\end{eqnarray}

The procedure described above has been applied in order
to generate non-renormalizable terms for down-type quarks of Eq.
(\ref{eq:qlag}). In this case,  we introduce VLQs $G_{iR},G_{iL}$,
where $i=1,\cdots,30$. Below we have given invariant Lagrangians in the
form of ${\cal L}^d_{ij}$, which generate the $h^d_{ij}$ term of Eq.
(\ref{eq:qlag}), after integrating the heavy VLQs and flavons. Since
these Lagrangians are invariant under the $U(1)_F\times Z_2$, the charges
of VLQs under this symmetry have been fixed in terms of corresponding
charges of quarks and flavons. These charges are given in Eq. (\ref{eq:u1fd}).
\begin{eqnarray}\label{eq:h33d-UV-Lag}
\mathcal{L}^d_{33}= \bar{Q}_{3L}\phi_1 G_{1R}+ F_1 \bar{G}_{1R}G_{1L}+X \bar{G}_{1L}G_{2R}+ F_1 \bar{G}_{2R}G_{2L}+ X \bar{G}_{2L}d_{3R}+h.c..
\end{eqnarray}
\begin{eqnarray}\label{eq:h32d-UV-Lag}
	\mathcal{L}^d_{32}&=& \bar{Q}_{3L}\phi_1 G_{1R}+ F_1 \bar{G}_{1R}G_{1L}+ X \bar{G}_{1L}G_{2R}+ F_1 \bar{G}_{2R}G_{2L}+  X \bar{G}_{2L}G_{3R}\nonumber \\
	&&+ M \bar{G}_{3R}G_{3L}	+ X \bar{G}_{3L}G_{4R}+ F_2^*\bar{G}_{4R}G_{4L}+ X \bar{G}_{4L}d_{2R}+ h.c..
\end{eqnarray}
\begin{eqnarray}\label{eq:h23d-UV-lag}
\mathcal{L}^d_{23} &= & \bar{Q}_{2L}\phi_1 G_{5R}+ F_1 \bar{G}_{5R}G_{5L}+X\bar{G}_{5L}G_{6R}+F_1 \bar{G}_{6R}G_{6L}+X \bar{G}_{6L}G_{7R}\nonumber \\	
&&+F_1 \bar{G}_{7R}G_{7L}+ X \bar{G}_{7L}G_{8R}+M \bar{G}_{8R}G_{8L}+\bar{G}_{8L}d_{3R}+h.c..
\end{eqnarray}
 \begin{eqnarray}\label{eq:u22d-UV-Lag}
 \mathcal{L}^d_{22}&=&  \bar{Q}_{2L}\phi_1 G_{5R}+ F_1 \bar{G}_{5R}G_{5L}+X\bar{G}_{5L}G_{6R}+F_1 \bar{G}_{6R}G_{6L}+X \bar{G}_{6L}G_{7R}\nonumber \\
 	&& +F_1 \bar{G}_{7R}G_{7L}+X\bar{G}_{7L}G_{9R}+F_2^* \bar{G}_{9R}G_{9L}+X\bar{G}_{9L}d_{2R}+h.c..
 \end{eqnarray}
\begin{eqnarray}\label{eq:h13d-UV-Lag}
\mathcal{L}^d_{13}&=& \bar{Q}_{1L}\phi_1 G_{10R}+F_1 \bar{G}_{10R}G_{10L}+X\bar{G}_{10L}G_{11R}+	+F_1 \bar{G}_{11R}G_{11L}+X\bar{G}_{11L}G_{12R} \nonumber \\
&&+F_1 \bar{G}_{12R}G_{12L}+X \bar{G}_{12L}G_{13R}+F_2 \bar{G}_{13R}G_{13L}+X \bar{G}_{13L}G_{14R}+F_2 \bar{G}_{14R}G_{14L}\nonumber \\
&&+X \bar{G}_{14L}G_{15R}+F_2 \bar{G}_{15R}G_{15L}+X\bar{G}_{15L}d_{3R}+h.c..
\end{eqnarray}
\begin{eqnarray}\label{eq:h12d-UV-Lag}
\mathcal{L}^d_{12}&=& \bar{Q}_{1L}\phi_1 G_{10R}+F_1 \bar{G}_{10R}G_{10L}+X\bar{G}_{10L}G_{11R}+F_1 \bar{G}_{11R}G_{11L}+X\bar{G}_{11L}G_{12R}\nonumber \\
&&+F_1 \bar{G}_{12R}G_{12L}+X \bar{G}_{12L}G_{13R}+F_2 \bar{G}_{13R}G_{13L}+X \bar{G}_{13L}G_{14R}+F_2 \bar{G}_{14R}G_{14L}\nonumber \\
&&X \bar{G}_{14L}G_{16R}+M \bar{G}_{16R}G_{16L}+X\bar{G}_{16L}d_{2R}+h.c..
\end{eqnarray}
\begin{eqnarray}\label{eq:h11d-UV-Lag}
\mathcal{L}^d_{11}&=&\bar{Q}_{1L}\phi_1 G_{10R}+F_1 \bar{G}_{10R}G_{10L}+X\bar{G}_{10L}G_{11R} 
 +F_1 \bar{G}_{11R}G_{11L}+X\bar{G}_{11L}G_{12R}\nonumber \\
 &&+F_1 \bar{G}_{12R}G_{12L}+X\bar{G}_{12L}G_{17R}+F_1 \bar{G}_{17R}G_{17L}+X\bar{G}_{17L}G_{18R} +F_1\bar{G}_{18R}G_{18L}\nonumber \\
 &&+X\bar{G}_{18L}G_{19R}+F_2^* \bar{G}_{19R}G_{19L}+X\bar{G}_{19L}d_{1R}+h.c..	
\end{eqnarray}
\begin{eqnarray}\label{eq:h21d-Uv-Lag}
\mathcal{L}^d_{21}&=& \bar{Q}_{2L}\phi_1 M_{5R}+ F_1 \bar{G}_{5R}G_{5L}+X\bar{G}_{5L}G_{6R}+F_1 \bar{G}_{6R}G_{6L}+X \bar{G}_{6L}G_{7R}\nonumber \\
 &&+F_1 \bar{G}_{7R}G_{7L} +X\bar{G}_{7L}G_{9R}+F_2^* \bar{G}_{9R}G_{9L} + X\bar{G}_{9L}G_{20R} + F_2^*\bar{G}_{20R}G_{20L} \nonumber \\
&& +X\bar{G}_{20L}G_{21R}+F_2^* \bar{G}_{21R}G_{21L}
+ X\bar{G}_{21L}G_{22R}+ F_2^* \bar{G}_{22R}G_{22L}+ X\bar{G}_{22L}G_{23R}\nonumber \\
&&+F_1 \bar{G}_{23R}G_{23L} + X \bar{G}_{23L}G_{24R} + F_1\bar{G}_{24R}G_{24L} + X \bar{G}_{24L}G_{25R}+M \bar{G}_{25R}G_{25L}\nonumber \\
&&+X \bar{G}_{25L}d_{1R}+h.c..
\end{eqnarray}
\begin{eqnarray}\label{eq:h31d-UV-Lag}
\mathcal{L}^d_{31}&=& \bar{Q}_{3L}\phi_1 G_{1R}+ F_1 \bar{G}_{1R}G_{1L}+ X \bar{G}_{1L}G_{2R}+ F_1 \bar{G}_{2R}G_{2L}+ X\bar{G}_{2L}G_{26R}\nonumber \\
&&+F_1\bar{G}_{26R}G_{26L} + X\bar{G}_{26L}G_{27R} + F_1 \bar{G}_{27R}G_{27L} + X\bar{G}_{27L}G_{28R} + F_2^*\bar{G}_{28R}G_{28L}\nonumber \\
&&+X\bar{G}_{28L}G_{29R}+F_2^* \bar{G}_{29R}G_{29L} + X\bar{G}_{29L}G_{30R} + F_2^*\bar{G}_{30R}G_{30L} +X\bar{G}_{30L}G_{31R}\nonumber \\
&&+F_2^*\bar{G}_{31R}G_{31L} + X \bar{G}_{31L}d_{1R}+h.c..	
\end{eqnarray}
\begin{eqnarray}\label{eq:u1fd}
U(1)_F	&:& G_{1R}\rightarrow q_{3f},\quad G_{1L},G_{2R}\rightarrow q_{3f}-f_1\quad G_{2L},G_{3R},G_{3L},G_{4R},G_{25R}\rightarrow q_{3f}-2f_1,\nonumber \\
	&& G_{4L}\rightarrow q_{3f}-2f_1+f_2,\quad G_{5R}\rightarrow q_{2f},\quad G_{5L},G_{6R}\rightarrow q_{2f}-f_1,\nonumber \\
	&& G_{6L},G_{7R}\rightarrow q_{2f}-2f_1,\quad G_{7L},G_{8R},G_{8L},G_{9R}\rightarrow q_{2f}-3 f_1\nonumber \\
	&& G_{9L},G_{20R}\rightarrow q_{2f}-3 f_1+f_2,\quad G_{10R}\rightarrow q_{1f},\quad G_{10L},G_{11R}\rightarrow q_{1f}-f_1,\nonumber \\
	&& G_{11L},G_{12R}\rightarrow q_{1f}-2 f_1,\quad G_{12L},G_{13R},G_{17R}\rightarrow q_{1f}-3 f_1,\nonumber \\ &&G_{13L},G_{14R}\rightarrow q_{1f}-3 f_1-f_3,\quad G_{14L},G_{15R},G_{16L},G_{16R}\rightarrow q_{1f}-3 f_1-2 f_2 \nonumber \\
	&& G_{15L}\rightarrow q_{1f}-3 f_1-3f_3,\quad G_{17L},G_{18R}\rightarrow q_{1f}-4 f_1,\quad G_{18L},G_{19R}\rightarrow q_{1f}-5 f_1,\nonumber \\ 
	&& G_{19L}\rightarrow q_{1f}-5 f_1-f_2,\quad G_{20L},G_{21R}\rightarrow q_{2f}-3 f_1+2 f_2,\nonumber \\
	&& G_{21L},G_{22R}\rightarrow q_{2f}-3f_1+3 f_2,\quad G_{22L},G_{23R}\rightarrow q_{2f}-3 f_1+4 f_2, \nonumber \\
	&& G_{23L},G_{24R}\rightarrow q_{2f}--4 f_1+4 f_2,\quad G_{24L},G_{25L},G_{25R}\rightarrow q_{2f}-5 f_1+4 f_2, \nonumber \\
	&& G_{26L},G_{27R}\rightarrow q_{3f}-3 f_1,\quad G_{27L},G_{28R}\rightarrow q_{3f}-4 f_1,\quad G_{28L}, G_{29R}\rightarrow q_{3f}-4 f_1+f_2,\nonumber \\
	&& G_{29L},G_{30R}\rightarrow q_{3f}-4 f_1+2 f_2,\quad G_{30L},G_{31R}\rightarrow q_{3f}-4 f_1+3 f_2,\nonumber \\
	&& G_{31L}\rightarrow q_{3f}-4 f_1+4 f_2.
\nonumber \\
Z_2 &:&	G_{1L},G_{1R}, G_{3L}, G_{3R}, G_{5L}, G_{5R}, G_{7L}, G_{7R}, G_{10L}, G_{10R},G_{12L}, G_{12R}, G_{14L}, G_{14R}, G_{18L},  \nonumber \\ 
&&G_{18R}, G_{20L}, G_{20R}, G_{22L}, G_{22R}, G_{24L}, G_{24R}, G_{26L}, G_{26R}, G_{28L}, G_{28R}, G_{30L}, G_{30R} \rightarrow \text{odd}. \nonumber \\
&&	 G_{2L},G_{2R}, G_{4L}, G_{4R}, G_{6L}, G_{6R}, G_{8L}, G_{8R}, G_{9L}, G_{9R}, G_{11L}, G_{11R}, G_{13L}, G_{13R},G_{15L},\nonumber \\
 && G_{15R}, G_{16L}, G_{16R}, G_{17L}, G_{17R}, G_{19L}, G_{19R}, G_{21L}, G_{21R}, G_{23L}, G_{23R}, G_{25L}, G_{25R},\nonumber \\
 && G_{27L}, G_{27R}, G_{29L}, G_{29R}, G_{31L}, G_{31R} \rightarrow \text{even}.
\end{eqnarray}

Since the Lagrangians of Eqs. (\ref{eq:h33d-UV-Lag}) -
(\ref{eq:h31d-UV-Lag}) are invariant under $U(1)_F$, nine relations exist
among the $U(1)_F$ charges of quarks and flavons. These relations can be
solved consistently along with Eq. (\ref{eq:urel}). After doing this,
the $U(1)_F$ charges of singlet down-type quarks can be expressed as
\begin{equation}
d_{3f}=q_{3f}-2f_1,\quad d_{2f}=q_{3f}-2f_1+f_2,\quad
d_{1f}=q_{3f}-4f_1+4f_2.
\end{equation}

In this section, we have described our model for quark sector and also the UV
completion to this model. As part of this whole construction, we have introduced extra
fields and symmetries into our model. We have summarized these fields and symmetries
in Tab. \ref{table:quark fields and roles}.
\begin{table}[h!]
	\begin{center}
		\begin{tabular}{|c|c|}
			\hline
			additional field &  role \\
			\hline
			$X$ & to generate hierarchy in quark masses and also $CP$ violation \\
			& in quark sector\\
			\hline
			$F_1,F_2$ & to generate masses for VLQs in the UV completion of
			our model\\
			\hline 
			$K_{iL},K_{iR}$($i=1,\cdots,16$) & to generate effective Yukawa
			couplings for up-type quarks \\
			& from UV completion our model \\
			\hline
			$G_{iL},G_{iR}$($i=1,\cdots,31$) & to generate effective Yukawa
			couplings for down-type quarks \\
			& from UV completion our model\\
			\hline
		\end{tabular}
	\end{center}
	\begin{center}
		\begin{tabular}{|c|c|}
		\hline
		additional symmetry & role \\
		\hline
		$U(1)_F$ & to generate invariant terms in the UV completion of our model\\
		\hline
		\end{tabular}
	\end{center}
	\caption{Additional fields and symmetry, along with their roles, in the
	quark sector of our model.}
	\label{table:quark fields and roles}
\end{table}

\section{Full scalar potential}

In Sec. 3, we have given the analysis of scalar potential for the model
described in Sec. 2. However, the model in Sec. 2 addresses problems related
to masses of leptons. Later, within the framework of the model of Sec. 2,
we have addressed hierarchy in the masses of quark fields in Sec. 5. While
addressing the hierarchy in quark masses, we have introduced additional
singlet scalar fields: $X,F_1,F_2$. These additional scalar fields can
give extra terms with the doublet and triplet Higgses in the scalar potential.
These extra terms may change the results derived in Sec. 3. For this purpose,
in this section, we give the full scalar potential of our model. After
minimizing the full scalar potential, we demonstrate that the above mentioned
singlet scalar fields do not change the main conclusions of the analysis
of Sec. 3. It is to remind that the following are the main conclusions of
Sec. 3: (i) triplet Higgs acquire real VEV, (ii) VEVs of doublet Higgses
$\Phi_{2,3}$ explain the hierarchy between $m_\mu$ and $m_\tau$.

The full scalar potential of our model is
\begin{eqnarray}
	V_{{\rm full}}=V_{inv}+ V_{X,F_1,F_2}+V_{\slashed{K}}+V^\prime_{\slashed{K}}
\end{eqnarray}
Here, $V_{X,F_1,F_2}$ is the invariant scalar potential of our model,
arising due to the singlet fields $X,F_1,F_2$. $V^\prime_{\slashed{K}}$ contain
potential terms due to $X,F_1,F_2$, which violate $K$-symmetry explicitly.
It is to remind here that minimization of $V_{inv}+V_{\slashed{K}}$ has been discussed
in Sec. 3. First we find a minimum after minimizing $V_{inv}+V_{X,F_1,F_2}$.
Later we study the shift in this minimum due to the presence of $K$-violating
terms. In this regard, the minimization of $V_{inv}$, after applying
the $Z_3$ symmetry, has been studied in Sec. 3.3. Now, let us see if this
minimization can be affected due to $V_{X,F_1,F_2}$. The form for this
potential is given below.
\begin{eqnarray}
	V_{X,F_1,F_2}&=&-m_X^2 (X^*X)-m_{F_1}^2(F_1^* F_1)-m_{F_2}^2(F_2^* F_2)+\lambda_X (X^*X)^2 +\lambda_{F_1}(F_1^* F_1)^2\nonumber \\
	&& +\lambda_{F_2}(F_2^* F_2)^2+ A(X^2+{X^*}^2)+B(X^4+{X^*}^4)+ \lambda_X^{\prime}(X^3X^{*}+{X^*}^3X) \nonumber \\
&&+\lambda_{F_1F_2}(F_1^* F_1)(F_2^* F_2) +\lambda_{F_1 X}(F_1^*F_1)(X^*X) +\lambda_{F_1 X}^{\prime}(F_1^*F_1)(X^2+{X^*}^2)\nonumber \\
&&+\lambda_{F_2 X}(F_2^*F_2)(X^*X)+\lambda_{F_2 X}^{\prime}(F_2^*F_2)(X^2+{X^*}^2)	+\lambda_{\phi_1 X}(\phi_1^{\dagger}\phi_1)(X^*X) \nonumber \\
	&& +\lambda_{\phi_1 X}^{\prime}(\phi_1^{\dagger}\phi_1)(X^2+{X^*}^2)+\lambda_{\phi_2 X}(\phi_2^{\dagger}\phi_2+\phi_3^{\dagger}\phi_3)(X^*X)\nonumber \\
	&&+\lambda_{\phi_2 X}^{\prime}(\phi_2^{\dagger}\phi_2+\phi_3^{\dagger}\phi_3)(X^2+{X^*}^2)+\lambda_{\Delta X}\rm Tr(\Delta^{\dagger}\Delta)(X^*X)\nonumber \\
	&& +\lambda_{\Delta X}^{\prime}{\rm Tr}(\Delta^{\dagger}\Delta)(X^2+{X^*}^2)+\lambda_{F_1\phi_1}(F_1^*F_1)(\phi_1^{\dagger}\phi_1)+\lambda_{F_2\phi_1}(F_2^*F_2)(\phi_1^{\dagger}\phi_1)\nonumber \\
	&& +\lambda_{F_1\phi_2}(F_1^*F_1)(\phi_2^{\dagger}\phi_2+\phi_3^{\dagger}\phi_3)+\lambda_{F_2\phi_2}(F_2^*F_2)(\phi_2^{\dagger}\phi_2+\phi_3^{\dagger}\phi_3)\nonumber \\
	&& +\lambda_{F_1\Delta}{\rm Tr}(\Delta^{\dagger}\Delta)(F_1^*F_1)+\lambda_{F_2\Delta}{\rm Tr}(\Delta^{\dagger}\Delta)(F_2^*F_2).
\label{eq:vxf}
	\end{eqnarray}
In the above equation, all the parameters are real due to hermiticity and
$CP$ symmetry.

In Eq. (\ref{eq:vxf}), $F_1$ and $F_2$ appear in the form of $F_1^*F_1$ and
$F_2^*F_2$, respectively. As a result of this, we can take the VEVs of
$F_1$ and $F_2$ to be real. Hence, we can parameterize the VEVs for
$X,F_1,F_2$ as
\begin{eqnarray}\label{eq:new VEV para}
	 \langle X \rangle =v_{X}e^{i\theta_{X}},\quad \langle F_1 \rangle =v_{f_1},\quad \langle F_2 \rangle =v_{f_2}.
\end{eqnarray}
Here, $\theta_X$ is the phase in the VEV of $X$.
After using the above VEVs and also Eq. (\ref{eq:para}) in Eq. (\ref{eq:vxf}),
we get
\begin{eqnarray}\label{eq:full potnetial VEVs}
\langle V_{X,F_1,F_2}\rangle&\ni& 2 v_X^2[ A +\lambda_X^{\prime} v_X^2 +\lambda_{\phi_1 X}^{\prime}v_1^2+\lambda_{\phi_2 X}^{\prime}v^2+\lambda_{\Delta X}{v^{\prime}}^2+\lambda_{F_1 X}^{\prime}v_{f_1}^2+
	+\lambda_{F_2 X}^{\prime}v_{f_2}^2]\cos2\theta_X \nonumber \\
&&	+2 B v_X^4 \cos4\theta_{X}.
\end{eqnarray}
In the above equation, we have not written constant terms which do not
contain phases of the VEVs of the fields.
From the above equation we can see that $\theta_X$ do not mix with the
phases in the VEVs of $\phi_{2,3}$ and $\Delta$. Hence,
the minimization of $\langle V_{inv}\rangle$, which is presented in Sec. 3.3,
is not affected due to $\langle V_{X,F_1,F_2}\rangle$. As a result of
this, $\Delta$ can acquire real VEV, and moreover, Eq. (\ref{eq:z3vac})
is still valid. Now, from the minimization of
$\langle V_{X,F_1,F_2}\rangle$ with respect to $\theta_X$, we get
\begin{eqnarray}
	\cos2\theta_{X}&=&-\frac{1}{4 B v_X^2}[A+\lambda_X^{\prime}v_X^2+\lambda_{\phi_1 X}^{\prime}v_1^2+\lambda_{\phi_2 X}^{\prime}v^2+\lambda_{\Delta X}^{\prime}{v^{\prime}}^2+\lambda_{F_1 X}^{\prime}v_{f_1}^2+\lambda_{F_2 X}^{\prime}v_{f_2}^2]
\nonumber \\
\label{eq:thx}
\end{eqnarray}
The above relation corresponds to the minimum for $\theta_X$, provided the
below condition is satisfied.
\begin{eqnarray}
16 B^2 v_X^4&>& [A+\lambda_X^{\prime}v_X^2+\lambda_{\phi_1 X}^{\prime}v_1^2+\lambda_{\phi_2 X}^{\prime}v^2+\lambda_{\Delta X}^{\prime}{v^{\prime}}^2+\lambda_{F_1 X}^{\prime}v_{f_1}^2+\lambda_{F_2 X}^{\prime}v_{f_2}^2]^2 .
\end{eqnarray}
Here we have shown that $X$ can acquire complex VEV. This is necessary
to achieve, in order to generate the $CP$ violation in quark sector, which
is discussed in Sec. 5.

After minimizing $\langle V_{inv}\rangle+\langle V_{X,F_1,F_2}\rangle$,
we have shown that the minimum can be given by Eqs. (\ref{eq:z3vac})
and (\ref{eq:thx}). Now, this minimum can be shifted by small amount
due to $K$-violating
terms. Terms in $V_{\slashed{K}}$ are presented in Sec. 3. Below we give the form
for $V_{\slashed{K}}^\prime$.
\begin{eqnarray}\label{eq:new soft}
	V_{\slashed{K}}^{\prime}&=& \delta\lambda_{\phi_2 X}(\phi_2^{\dagger}\phi_2)(X^*X)+\delta\lambda_{\phi_2 X}^{\prime}(\phi_3^{\dagger}\phi_3)(X^*X)+\delta\lambda_{\phi X}(\phi_2^{\dagger}\phi_2)(X^2+{X^*}^2)\nonumber \\
	&&+\delta\lambda_{\phi X}^{\prime}(\phi_3^{\dagger}\phi_3)(X^2+{X^*}^2)+\delta \lambda_{\phi_2 F_1}(\phi_2^{\dagger}\phi_2)(F_1^{*}F_1)+\delta \lambda_{\phi_2 F_1}^{\prime}(\phi_3^{\dagger}\phi_3)(F_1^{*}F_1)\nonumber \\
	&& +\delta \lambda_{\phi_2 F_2}(\phi_2^{\dagger}\phi_2)(F_2^{*}F_2)+\delta \lambda_{\phi_2 F_2}^{\prime}(\phi_3^{\dagger}\phi_3)(F_2^{*}F_2)+i\delta\lambda_{21}(\phi_2^{\dagger}\phi_3-\phi_3^{\dagger}\phi_2)(X^*X)\nonumber \\
	&& +i\delta\lambda_{22}(\phi_2^{\dagger}\phi_3-\phi_3^{\dagger}\phi_2)(X^2+{X^*}^2)+i\delta\lambda_{23}(\phi_2^{\dagger}\phi_3-\phi_3^{\dagger}\phi_2)(F_1^*F_1)\nonumber \\
&&+i\delta\lambda_{24}(\phi_2^{\dagger}\phi_3-\phi_3^{\dagger}\phi_2)(F_2^*F_2).
\end{eqnarray}
All the parameters in the above equation are real due to either $CP$ symmetry
or hermiticity of the potential. These parameters should be small as compared
to the parameters in $V_{inv}+V_{X,F_1,F_2}$, since the above
potential violates $K$ symmetry by a small amount. We can see that the VEVs of
$X,F_1,F_2$ in $V_{\slashed{K}}^\prime$ can give additional contribution to
$f_0$ and $f_\zeta$ of Eq. (\ref{eq:Z3-f0-fzeta}). Since the parameters of
$V_{\slashed{K}}^\prime$
are small, we can notice that the contribution due to $X,F_1,F_2$ can
be of the same order of the terms which are already obtained for
$f_0$ and $f_\zeta$ in Eq. (\ref{eq:Z3-f0-fzeta}). As a result of this,
the hierarchy
in $m_\mu$ and $m_\tau$ can be explained in our framework.

\section{Phenomenology of our model}

In Secs. 3 and 6 we have analyzed the minimum of the scalar potential of
our model. One needs to study if this minimum corresponds to global or
local minimum. Following the studies made in Refs. \cite{xu}, we expect
some additional conditions to be imposed on the parameters of our model
in order for the minimum of the potential in this work to be global. We
work on the vacuum stability of our scalar potential in future.

The scalar fields, which are proposed in our model, are: three Higgs doublets,
one Higgs triplet and three singlet scalar fields. We can choose the
$U(1)_F$ symmetry of our model be gauged. As a result of this, after
electroweak symmetry breaking,
the following fields remain in the theory: one doubly charged scalar,
three singly charged scalars, seven neutral scalars and five pseudo scalars.
In the case I, which is described in Sec. 3, the scalar fields belonging
to the triplet Higgs can have masses around $10^{12}$ GeV. But otherwise,
we can choose the parameters in the scalar potential of our model in such
a way that all the scalar fields can have masses less than or about 1 TeV.
In the case where the masses for the scalar fields are less than 1 TeV,
one can study the collider phenomenology. For this study, one needs to know the
interaction of the scalar fields with the standard model particles. We can
see that the scalar components of Higgs doublets and Higgs triplet have
gauge interactions. For the field $\phi_1$, it has Yukawa interactions with
quark fields. Hence, the scalars belonging to doublet and triplet Higgses
can be produced at the LHC experiment either via gauge or strong interactions.
After production, subsequently they will decay into standard model fields.
In the case of singlet scalars $X,F_1,F_2$, the flavons have Yukawa
interactions with VLQs. Moreover, these flavons interact with doublet
and triplet Higgses in scalar potential. As for the $X$ field, it has
Yukawa interactions containing a VLQ and a right-handed quark field. Moreover,
$X$ has interactions with Higgs fields in the scalar potential. Since VLQs
are color triplets, they can be produced at the LHC experiment via strong
interactions. From the decay of these VLQs, one can produce the above
mentioned singlet scalars in the LHC experiment. Studying
the collider phenomenology of this model is the beyond
the scope of this work.

In the lepton sector of our model, the Yukawa couplings for charged leptons
are diagonal. Hence, these Yukawa interactions are flavor conserving.
On the other hand, the Yukawa couplings for neutrinos are flavor violating.
As a result of this, the singly and doubly charged triplet Higgs fields can
drive flavor violating decays of the form $\ell\to3\ell^\prime$ and
$\ell\to\ell^\prime\gamma$. However, it is stated in Sec. 2 that the Yukawa
couplings for neutrinos are suppressed by about $10^{-3}$. Hence, the
branching ratios for the above mentioned decays are suppressed even if the
components of triplet Higgs can have masses around few hundred GeV. As a
result of this, constraints due to non-observation of charged lepton flavor
violating decays \cite{pdg} are satisfied in our model.

The phenomenology of our model in quark sector is similar to that discussed
in Ref. \cite{lmn}. In this regard, the $X$ field can cause flavor
changing neutral currents at tree level in our model. As a result of this,
there can
be mass splitting in the $K^0-\bar{K}^0$ and $D^0-\bar{D}^0$ due to the
mediation of $X$. We have estimated the above mentioned mass splittings,
using the procedures described in Refs. \cite{bn,lmn}. For this
purpose, we define $\beta=\frac{v_1}{M}$. In our calculations we have
taken $\epsilon\sim\beta=\frac{1}{5.5}$ and the mass of $X$ as 1 TeV.
The Yukawa couplings for $h^{d,u}_{12}$ are given in Eq. (\ref{eq:numer}).
We have chosen $h_{21}^d\sim1$ and $h_{21}^u=-0.7$. Using the above set of
parameters,
we have found $\Delta m_{K}\approx 10^{-16}$ GeV and
$\Delta m_D\approx 10^{-15}$ GeV. These numerical values are smaller than
the corresponding current experimental values, which are as follows:
$\Delta m_{K}=3.5 \times 10^{-15}$ GeV and
$\Delta m_D=2.35 \times 10^{-14}$ GeV \cite{pdg}. Hence, our model satisfies
the constraints due to the mass splitting in $K^0-\bar{K}^0$ and
$D^0-\bar{D}^0$.

By choosing $U(1)_F$ be gauged, the gauge boson corresponding to this
symmetry, $Z^\prime$, can be massive. The mixing between $Z-Z^\prime$
is constrained to be very small. In this regard, phenomenology due to
$Z^\prime$ can be studied in our model. To know some phenomenology on
$Z^\prime$, see Refs. \cite{Grossmann:2010ea,Erler:2011ud}.

\section{Conclusions}

In this work, we have proposed a model which explains the maximal values
for $\theta_{23}$ and $\delta_{CP}$ in the lepton sector. To achieve
this purpose we have introduced three Higgs doublets and one Higgs triplet.
This model is based
on $\mu-\tau$ reflection symmetry and type II seesaw mechanism. In this
model, to explain the above observables, the VEV of triplet Higgs should be
real. Moreover, due to $\mu-\tau$ reflection symmetry, the masses for muon and
tau can be of the same order. After introducing the $K$ symmetry and explicit
violation of it by a small amount, we have studied the minimization of scalar
potential of our model. Thereafter, we have shown that the VEV of triplet Higgs
can be real, apart from explaining the hierarchy in the muon and tau masses.
In addition to predicting the above observables, the mass matrix for
neutrinos in our model can also predict about neutrino mass ordering and
smallness of
$\theta_{13}$, if the elements of this matrix satisfy certain conditions.
These conditions are given in Sec. 4. To explain these conditions, one
has to propose a new mechanism in addition to $CP$ symmetry. Although
we do not have a mechanism to explain all the conditions given in Sec. 4,
we have attempted to give one mechanism to explain condition (i) for the
case of NO. This mechanism is presented in the Appendix.

Since in our model three Higgs doublets exist, we
have studied the Yukawa couplings between quarks and these doublets by proposing
$CP$ transformations for quark fields. After employing a certain texture for
these Yukawa couplings, we have consistently explained the quark masses and
mixing pattern. To employ this texture in the quark sector of our model, we
have introduced additional fields like VLQs and singlet scalars. One of
these singlet scalars should acquire complex VEV in order to generate the
$CP$ violating phase in quark sector. Finally, we have analyzed the
scalar potential containing the singlet scalars and the above mentioned
Higgs fields. After this analysis, we have demonstrated that masses and mixing
pattern in lepton and quark sectors can be consistently explained.

\section*{Appendix: A model for achieving condition (i) in the case of NO}

In Sec. 4, we have described some conditions on the elements of neutrino
mass matrix, which can predict the case of NO or IO and also about the
smallness of $\theta_{13}$. These conditions are purely phenomenological
and cannot be achieved with just
the $CP$ symmetry. Additional mechanism should be proposed in order
to satisfy these conditions. For the case of NO, condition (i) can
be achieved if we propose an extra $U(1)_S$ symmetry and the
singlet scalar fields $S_1,S_2$. Under the $U(1)_S$ symmetry, we
consider the following charge assignments, where $l$ is some
non-zero rational number:$D_{eL}\to l$, $S_1\to-2l$, $S_2\to-l$.
Under $U(1)_S$, $e_R$ should transform like $D_{eL}$ and rest of
the fields in our model are singlets. $S_1,S_2$ transform under the
$CP$ symmetry as $S_{1,2}\to S_{1,2}^*$. With the above charge
assignments, the Yukawa terms for $D_{eL}$ in Eq. (\ref{eq:lagnu})
are forbidden. Now, these terms can be effectively generated by the
following invariant terms.
\begin{equation}
	Y_e\frac{S_1}{M}\bar{D}^c_{eL}i\sigma_2\Delta D_{eL}
	+Y_\mu\frac{S_2}{M}\bar{D}^c_{eL}i\sigma_2\Delta D_{\mu L}
	+Y_\tau\frac{S_2}{M}\bar{D}^c_{eL}i\sigma_2\Delta D_{\tau L}+h.c..
	\label{eq:nonre}
\end{equation}
Here, $M$ is a mass scale which is analogous to that in the quark sector
Lagrangian of Eq. (\ref{eq:qlag}). The above non-renormalizable terms can
be generated by studying the UV completion for these terms, where one can
propose heavy vector-like leptons whose masses are around $M$. The process of
this UV completion is analogous to what we describe in Sec. 5.2.
In order for Eq. (\ref{eq:nonre}) to be invariant under $CP$ symmetry, $Y_e$
should be real and $Y_\mu=Y_\tau^*$. After $U(1)_S$ symmetry
is spontaneously broken, terms in Eq. (\ref{eq:nonre}) effectively generate the
Yukawa couplings $Y^\nu_{ee}, Y^\nu_{e\mu},Y^\nu_{e\tau}$.
Moreover, by taking
$\frac{\langle S_{1,2}\rangle}{M}\sim0.1$, condition (i) for the case of NO
is satisfied. Since $Y^\nu_{ee}$ is real and
$Y^\nu_{e\mu}=(Y^\nu_{e\tau})^*$, $\langle S_{1}\rangle$ and
$\langle S_{2}\rangle$ should be real. We justify this statement by studying the
scalar potential for these fields.

The scalar potential, which is invariant under $CP\times Z_2\times Z_3\times
 K \times U(1)_F\times U(1)_S$ and containing $S_{1,2}$ can be written as
\begin{eqnarray}
V_{S_1,S_2}&=&-m_{S_1}^2 (S_1^* S_1)-	m_{S_2}^2 (S_2^* S_2)+\lambda_{S_1} (S_1^* S_1)^2+\lambda_{S_2} (S_2^* S_2)^2 + \lambda_{S_1 S_2} (S_1^* S_1)(S_2^* S_2) \nonumber \\
&& +\lambda_{\phi_1 S_1}(\phi_1^* \phi_1)(S_1^*S_1) + \lambda_{\phi_2 S_1}(\phi_2^* \phi_2+\phi_3^* \phi_3)(S_1^*S_1)+\lambda_{\phi_1 S_2}(\phi_1^* \phi_1)(S_2^*S_2)\nonumber \\
&& +\lambda_{\phi_2 S_2}(\phi_2^* \phi_2+\phi_3^* \phi_3)(S_2^*S_2)+\lambda_{\Delta S_1}{\rm Tr}(\Delta^{\dagger}\Delta)(S_1^* S_1)+\lambda_{\Delta S_2}{\rm Tr}(\Delta^{\dagger}\Delta)(S_2^* S_2)\nonumber  \\
	&& +\lambda_{S_1 X}(S_1^*S_1)(X^*X)+ \lambda_{S_2 X}(S_2^*S_2)(X^*X)+\lambda_{S_1 X}^{\prime}(S_1^*S_1)(X^2+{X^*}^2)\nonumber \\
	&& +\lambda_{S_2 X}^{\prime}(S_2^*S_2)(X^2+{X^*}^2)+\lambda_{F_1S_1}(F_1^*F_1)(S_1^*S_1)+\lambda_{F_1S_2}(F_1^*F_1)(S_2^*S_2).\nonumber \\
	&& +\lambda_{F_2S_1}(F_2^*F_2)(S_1^*S_1)+	+\lambda_{F_2S_2}(F_2^*F_2)(S_2^*S_2) + a (S_1^*S_2^2+S_1 {S_2^*}^2).
\end{eqnarray}
Now, the $K$-violating terms containing $S_{1,2}$ can be written as
\begin{eqnarray}
        V_{\slashed{K}}^{\prime\prime}&=& \delta \lambda_{\phi_2 S_1}(\phi_2^{\dagger}\phi_2)(S_1^{*}S_1)+\delta\lambda_{\phi_2 S_1}^{\prime}(\phi_3^{\dagger}\phi_3)(S_1^{*}S_1)+\delta\lambda_{\phi_2 S_2}(\phi_2^{\dagger}\phi_2)(S_2^*S_2)\nonumber \\
        && +\lambda_{\phi_2 S_2}^{\prime}(\phi_3^{\dagger}\phi_3)(S_2^*S_2)+i\delta\lambda_{25}(\phi_2^{\dagger}\phi_3-\phi_3^{\dagger}\phi_2)(S_1^*S_1)\nonumber \\
                &&+i\delta\lambda_{26}(\phi_2^{\dagger}\phi_3-\phi_3^{\dagger}\phi_2)(S_2^*S_2).
\end{eqnarray}
In the above two potentials, all the parameters are real due to hermiticity
or $CP$ symmetry.
Analogous to what we describe in Sec. 6, we can see that the potential
terms in $V_{S_1,S_2}$ and $V_{\slashed{K}}^{\prime\prime}$ do not alter the main
conclusions of Sec. 3.3. This means, even with the fields $S_{1,2}$,
$\Delta$ acquires real VEV, the VEVs of $\phi_{2,3}$ explain the
hierarchy in $m_\mu$ and $m_\tau$.

Here we demonstrate that $S_{1,2}$ can acquire real VEVs. In this regard,
we can see that the last term of $V_{S_1,S_2}$ can only contain the phases
in the VEVs of $S_{1,2}$. Hence, after parameterizing $\langle S_1\rangle
=v_{s_1}e^{\theta_{s_1}}$ and $\langle S_1\rangle=v_{s_2}e^{\theta_{s_2}}$,
we get
\begin{eqnarray}\label{eq:S1-S2-potential}
\langle V_{S_1,S_2}\rangle &\ni& + 2 a v_{s_1}v_{s_2}^2\cos(2\theta_{s_2}-\theta_{s_1})
\end{eqnarray}
The above term has a minimum at $2\theta_{s_2}-\theta_{s_1}=0$ when
$av_{s_1}<0$. To satisfy this minimum we can choose
$\theta_{s_1}=\theta_{s_2}=0$.
Now, the minimum at $\theta_{s_1}=\theta_{s_2}=0$ cannot be shifted by the
terms of $V_{\slashed{K}}^{\prime\prime}$, since $S_{1,2}$ appear in the form of
$S_1^*S_1$ and $S_2^*S_2$ in $V_{\slashed{K}}^{\prime\prime}$.
Hence, there exist a parameter region where the VEVs of $S_1$ and $S_2$
are real in this model.


\section*{Acknowledgement}

JG acknowledges Dr. Anirban Karan for some useful discussion.


\begin{thebibliography}{}

\bibitem{glo-fit}
P.~F.~de Salas, D.~V.~Forero, S.~Gariazzo, P.~Mart\'\i{}nez-Mirav\'e, O.~Mena, C.~A.~Ternes, M.~T\'ortola and J.~W.~F.~Valle,
JHEP \textbf{02}, 071 (2021)
doi:10.1007/JHEP02(2021)071
[arXiv:2006.11237 [hep-ph]].

\bibitem{tbm}
P.~F.~Harrison, D.~H.~Perkins and W.~G.~Scott,
Phys. Lett. B \textbf{530}, 167 (2002)
doi:10.1016/S0370-2693(02)01336-9
[arXiv:hep-ph/0202074 [hep-ph]];
P.~F.~Harrison and W.~G.~Scott,
Phys. Lett. B \textbf{535}, 163-169 (2002)
doi:10.1016/S0370-2693(02)01753-7
[arXiv:hep-ph/0203209 [hep-ph]];
Z.~z.~Xing,
Phys. Lett. B \textbf{533}, 85-93 (2002)
doi:10.1016/S0370-2693(02)01649-0
[arXiv:hep-ph/0204049 [hep-ph]].

\bibitem{hs}
P.~F.~Harrison and W.~G.~Scott,
Phys. Lett. B \textbf{547}, 219-228 (2002)
doi:10.1016/S0370-2693(02)02772-7
[arXiv:hep-ph/0210197 [hep-ph]].

\bibitem{rev}
Z.~z.~Xing and Z.~h.~Zhao,
Rept. Prog. Phys. \textbf{79}, no.7, 076201 (2016)
doi:10.1088/0034-4885/79/7/076201
[arXiv:1512.04207 [hep-ph]].

\bibitem{mutau}
M.~Holthausen, M.~Lindner and M.~A.~Schmidt,
JHEP \textbf{04}, 122 (2013)
doi:10.1007/JHEP04(2013)122
[arXiv:1211.6953 [hep-ph]];
F.~Feruglio, C.~Hagedorn and R.~Ziegler,
Eur. Phys. J. C \textbf{74}, 2753 (2014)
doi:10.1140/epjc/s10052-014-2753-2
[arXiv:1303.7178 [hep-ph]];
H.~J.~He, W.~Rodejohann and X.~J.~Xu,
Phys. Lett. B \textbf{751}, 586-594 (2015)
doi:10.1016/j.physletb.2015.10.066
[arXiv:1507.03541 [hep-ph]];
A.~S.~Joshipura and K.~M.~Patel,
Phys. Lett. B \textbf{749}, 159-166 (2015)
doi:10.1016/j.physletb.2015.07.062
[arXiv:1507.01235 [hep-ph]];
R.~N.~Mohapatra and C.~C.~Nishi,
JHEP \textbf{08}, 092 (2015)
doi:10.1007/JHEP08(2015)092
[arXiv:1506.06788 [hep-ph]];
S.~F.~King and C.~C.~Nishi,
Phys. Lett. B \textbf{785}, 391-398 (2018)
doi:10.1016/j.physletb.2018.08.056
[arXiv:1807.00023 [hep-ph]].

\bibitem{gl}
W.~Grimus and L.~Lavoura,
Phys. Lett. B \textbf{579}, 113-122 (2004)
doi:10.1016/j.physletb.2003.10.075
[arXiv:hep-ph/0305309 [hep-ph]].

\bibitem{bmv}
K.~S.~Babu, E.~Ma and J.~W.~F.~Valle,
Phys. Lett. B \textbf{552}, 207-213 (2003)
doi:10.1016/S0370-2693(02)03153-2
[arXiv:hep-ph/0206292 [hep-ph]].

\bibitem{t1s}
R.~N.~Mohapatra and G.~Senjanovic,
Phys. Rev. Lett. \textbf{44}, 912 (1980)
doi:10.1103/PhysRevLett.44.912;
J.~Schechter and J.~W.~F.~Valle,
Phys. Rev. D \textbf{22}, 2227 (1980)
doi:10.1103/PhysRevD.22.2227.

\bibitem{gl2}
W.~Grimus and L.~Lavoura,
J. Phys. G \textbf{30}, 73-82 (2004)
doi:10.1088/0954-3899/30/2/007
[arXiv:hep-ph/0309050 [hep-ph]].

\bibitem{pdg}
P.~A.~Zyla \textit{et al.} [Particle Data Group],
PTEP \textbf{2020}, no.8, 083C01 (2020)
doi:10.1093/ptep/ptaa104.

\bibitem{t2s}
M.~Magg and C.~Wetterich,
Phys. Lett. B \textbf{94}, 61-64 (1980)
doi:10.1016/0370-2693(80)90825-4;
J.~Schechter and J.~W.~F.~Valle,
Phys. Rev. D \textbf{22}, 2227 (1980)
doi:10.1103/PhysRevD.22.2227;
R.~N.~Mohapatra and G.~Senjanovic,
Phys. Rev. D \textbf{23}, 165 (1981)
doi:10.1103/PhysRevD.23.165;
G.~Lazarides, Q.~Shafi and C.~Wetterich,
Nucl. Phys. B \textbf{181}, 287-300 (1981)
doi:10.1016/0550-3213(81)90354-0.

\bibitem{gl3}
P.~M.~Ferreira, W.~Grimus, L.~Lavoura and P.~O.~Ludl,
JHEP \textbf{09}, 128 (2012)
doi:10.1007/JHEP09(2012)128
[arXiv:1206.7072 [hep-ph]].

\bibitem{maho-nis}
R.~N.~Mohapatra and C.~C.~Nishi,
Phys. Rev. D \textbf{86}, 073007 (2012)
doi:10.1103/PhysRevD.86.073007
[arXiv:1208.2875 [hep-ph]].

\bibitem{prwo}
R.~S.~Hundi and I.~Sethi,
Phys. Rev. D \textbf{102}, no.5, 055007 (2020)
doi:10.1103/PhysRevD.102.055007
[arXiv:2003.09809 [hep-ph]];
J.~Ganguly and R.~S.~Hundi,
Phys. Rev. D \textbf{103}, no.3, 035007 (2021)
doi:10.1103/PhysRevD.103.035007
[arXiv:2005.04023 [hep-ph]].

\bibitem{rasin}
A.~Rasin,
[arXiv:hep-ph/9708216 [hep-ph]];
A.~Rasin,
Phys. Rev. D \textbf{58}, 096012 (1998)
doi:10.1103/PhysRevD.58.096012
[arXiv:hep-ph/9802356 [hep-ph]].

\bibitem{bn}
K.~S.~Babu and S.~Nandi,
Phys. Rev. D \textbf{62}, 033002 (2000)
doi:10.1103/PhysRevD.62.033002
[arXiv:hep-ph/9907213 [hep-ph]].

\bibitem{lmn}
J.~D.~Lykken, Z.~Murdock and S.~Nandi,
Phys. Rev. D \textbf{79}, 075014 (2009)
doi:10.1103/PhysRevD.79.075014
[arXiv:0812.1826 [hep-ph]].

\bibitem{owql}
C.~C.~Li, J.~N.~Lu and G.~J.~Ding,
JHEP \textbf{02}, 038 (2018)
doi:10.1007/JHEP02(2018)038
[arXiv:1706.04576 [hep-ph]];
J.~N.~Lu and G.~J.~Ding,
Phys. Rev. D \textbf{98}, no.5, 055011 (2018)
doi:10.1103/PhysRevD.98.055011
[arXiv:1806.02301 [hep-ph]].

\bibitem{rr}
M.~H.~Rahat, P.~Ramond and B.~Xu,
Phys. Rev. D \textbf{98}, no.5, 055030 (2018)
doi:10.1103/PhysRevD.98.055030
[arXiv:1805.10684 [hep-ph]];
M.~J.~P\'erez, M.~H.~Rahat, P.~Ramond, A.~J.~Stuart and B.~Xu,
Phys. Rev. D \textbf{100}, no.7, 075008 (2019)
doi:10.1103/PhysRevD.100.075008
[arXiv:1907.10698 [hep-ph]];
M.~J.~P\'erez, M.~H.~Rahat, P.~Ramond, A.~J.~Stuart and B.~Xu,
Phys. Rev. D \textbf{101}, no.7, 075018 (2020)
doi:10.1103/PhysRevD.101.075018
[arXiv:2001.04019 [hep-ph]].

\bibitem{maet}
E.~Ma, M.~Raidal and U.~Sarkar,
Nucl. Phys. B \textbf{615}, 313-330 (2001)
[arXiv:hep-ph/0012101 [hep-ph]].

\bibitem{susy}
M. Drees, R. Godbole, and P. Roy, {\it Theory and Phenomenology of Sparticles}
(World Scientific, Singapore, 2004);
P. Bi\'{n}etruy, {\it Supersymmetry} (Oxford University Press, Oxford,
England, 2006);
S.~P.~Martin,
Adv. Ser. Direct. High Energy Phys. \textbf{21}, 1-153 (2010)
[arXiv:hep-ph/9709356 [hep-ph]].

\bibitem{cosmo}
N. Aghanim {\it et al.} (Planck Collaboration), arXiv:1807.06209 [astro-ph.CO].

\bibitem{gh}
D.~K.~Ghosh and R.~S.~Hundi,
Phys. Rev. D \textbf{85}, 013005 (2012)
doi:10.1103/PhysRevD.85.013005
[arXiv:1108.3428 [hep-ph]].

\bibitem{xu}
X.~J.~Xu,
Phys. Rev. D \textbf{94}, no.11, 115025 (2016)
doi:10.1103/PhysRevD.94.115025
[arXiv:1612.04950 [hep-ph]];
Phys. Rev. D \textbf{95}, no.11, 115019 (2017)
doi:10.1103/PhysRevD.95.115019
[arXiv:1705.08965 [hep-ph]].

	\bibitem{Grossmann:2010ea}
B.~N.~Grossmann, Z.~Murdock and S.~Nandi,
[arXiv:1011.5256 [hep-ph]].
\bibitem{Erler:2011ud}
J.~Erler, P.~Langacker, S.~Munir and E.~Rojas,
JHEP \textbf{11} (2011), 076
doi:10.1007/JHEP11(2011)076
[arXiv:1103.2659 [hep-ph]].
\end{thebibliography}
\end{document}